\documentclass{article}
\usepackage{graphicx, caption, amsmath}
  \usepackage{setspace}
\usepackage[margin=1in]{geometry}
\usepackage{authblk}
\DeclareCaptionLabelFormat{nospace}{#1#2}

\begin{document}
\title{multi-GPA-Tree: Statistical Approach for Pleiotropy Informed and Functional Annotation Tree Guided Prioritization of GWAS Results}
\author[1]{\small Aastha Khatiwada}
\author[2]{\small Ayse Selen Yilmaz}
\author[3]{\small Bethany J. Wolf}
\author[2]{\small Maciej Pietrzak}
\author[2,4,*]{\small Dongjun Chung}

\affil[1]{\footnotesize Department of Biostatistics and Bioinformatics, National Jewish Health, Denver, CO, USA}
\affil[2]{\footnotesize Department of Biomedical Informatics, The Ohio State University, Columbus, Ohio, USA}
\affil[3]{\footnotesize Department of Public Health Sciences, Medical University of South Carolina, Charleston, SC, USA}
\affil[4]{\footnotesize Pelotonia Institute for Immuno-Oncology, The James Comprehensive Cancer Center, The Ohio State University}
\affil[*]{\footnotesize To whom correspondence should be addressed (chung.911@osu.edu).}

\vspace{2ex}
%$^{\text{*}}$To whom correspondence should be addressed (chung.911@osu.edu).\\}}}

\date{}

\onehalfspacing
\normalsize
\maketitle
\abstract{
Genome-wide association studies (GWAS) have successfully identified over two hundred thousand genotype-trait associations. Yet some challenges remain. First, complex traits are often associated with many single nucleotide polymorphisms (SNPs), most with small or moderate effect sizes, making them difficult to detect. Second, many complex traits share a common genetic basis due to `pleiotropy' and and though few methods consider it, leveraging pleiotropy can improve statistical power to detect genotype-trait associations with weaker effect sizes. Third, currently available statistical methods are limited in explaining the functional mechanisms through which genetic variants are associated with specific or multiple traits. We propose multi-GPA-Tree to address these challenges. The multi-GPA-Tree approach can identify risk SNPs associated with single as well as multiple traits while also identifying the combinations of functional annotations that can explain the mechanisms through which risk-associated SNPs are linked with the traits.

First, we implemented simulation studies to evaluate the proposed multi-GPA-Tree method and compared its performance with an existing statistical approach.The results indicate that multi-GPA-Tree outperforms the existing statistical approach in detecting risk-associated SNPs for multiple traits. Second, we applied multi-GPA-Tree to a systemic lupus erythematosus (SLE) and rheumatoid arthritis (RA), and to a Crohn's disease (CD) and ulcertive colitis (UC) GWAS, and functional annotation data including GenoSkyline and GenoSkylinePlus. Our results demonstrate that multi-GPA-Tree can be a powerful tool that improves association mapping while facilitating understanding of the underlying genetic architecture of complex traits and potential mechanisms linking risk-associated SNPs with complex traits.\\
\textbf{Availability:} The multiGPATree software is available at https://dongjunchung.github.io/multiGPATree/.\\
}

\section{Introduction}
Increasing interest in identifying genomic regions associated with complex traits has resulted in a substantial increase in the number of reported GWAS studies and genotype-trait associations (https://www.ebi.ac.uk/gwas/) \cite{buniello2019nhgri}. Identification of previously unknown genotype-trait associations has improved estimation of heritability (genetic variation within a trait) for many complex traits. However, two major challenges persist. First, some proportion of heritability remains missing for many traits due to unidentified genotype-trait associations \cite{manolio1, leesh, maherb}. Polygenicity, a phenomenon that causes genetic variants to be associated with traits with weak or moderate effect sizes \cite{nikpay, priceal} is a plausible explanation for missing heritibility. The impact of polygenicity can theoretically be reduced by recruiting a larger GWAS sample size to increase statistical power to detect weak and moderate associations; however, large sample recruitment often requires more resources and is not always feasible due to limited trait prevalence in the population \cite{kundaje2015integrative}. An alternative to increasing sample size to improve statistical power is to exploit the pleiotropic relationship (shared genetic basis) between two or more traits by simultaneously integrating GWAS association summary statistics for multiple traits \cite{Andreassen, stearns, chung2017graph}. GWAS summary statistics are readily available to use as input through public data repositories \cite{buniello2019nhgri, dbgap} and are good proxy to using individual-level genotype-phenotype data that are harder to obtain. Second, majority of the GWAS identified loci are located in the non-coding regions \cite{giral2018into}, making it difficult to understand the functional mechanisms related to identified genotype-trait associations. For example, in autoimmune diseases, about $90\%$ of the causal genetic variants lie in non-coding regions, a bulk of which are located in regulatory DNA regions \cite{maurano2012systematic, farh2015genetic}. As such, utilizing genomic functional annotation information that can provide information related to different types of histone modifications, epigenetic and cell- and tissue-specific changes, etc. can be useful to decode the functional mechanisms linking risk-associated genetic variants to traits \cite{Ming2018, Zablocki2014, Schork2013a}. Therefore, integrative analysis of genetic data with genomic functional annotation data is a promising direction. 

Statistical methods built on the foundation of data integration approaches not only utilize information that are readily available in public data repositories but also overcome the challenges posed by polygenicity while simultaneously providing insights about underlying functional mechanisms related to one or more traits. Therefore, they are more advantageous and efficient. In recognizing the potential to enhance statistical power to detect associations through data integration approaches, several statistical methods focused on GWAS summary statistics have been developed \cite{Ming2018, Zablocki2014, Andreassen, chung2017graph, chung2014gpa, Ming439133}. These methods can broadly be classified into three distinct categories. 

The first category of methods exploit the pleiotropic relationship between two or more distinct traits by simultaneously integrating multiple GWAS association $p$-values together. Two favored methods in this category are the pleiotropy-informed conditional FDR approach \cite{Andreassen} and the graph-GPA approach \cite{chung2017graph}. The unifying goal of the two methods is to improve statistical power to prioritize one or more trait risk-associated SNPs. The conditional FDR approach shows improved detection of risk-associated SNPs for two psychiatric disorders, schizophrenia and bipolar disorder. Despite easy implementation of this approach, the lack of a model-based framework in estimating conditional FDR compromises the power to detect non-null associations and also to infer the properties of the non-null distribution. Moreover, this approach can only integrate a small number of GWAS traits. In contrast, graph-GPA can integrate large number of GWAS traits using a hidden Markov random field framework and its usefulness is demonstrated by integrating 12 traits (five psychiatric disorders, three autoimmune traits, two lipid-related traits and two cardiovascular traits) where clinically related traits form closely connected clusters. However, both methods fail to inform about functional relevance of risk-associated SNPs due to their inability to integrate functional annotations in their application.

The second category of methods integrate individual GWAS data with genotype-related functional annotation data. Two cutting-edge approaches in this category include the latent sparse mixed model (LSMM) approach \cite{Ming2018} and the covariate modulated false discovery rate (cmFDR) approach \cite{Zablocki2014}. In LSMM, functional annotations are integrated using a logistic mixed effects model framework where genic- and cell-type specific functional annotations are assumed to respectively have fixed and random effects and a sparse structure is imposed on the random effects to adaptively select cell-type specific functional annotations that may be relevant to a trait etiology. Through application of LSMM, Ming et al. discovered substantial enrichment of blood-related cell-type specific annotations for autoimmune diseases like systemic lupus erythematosus, rheumatoid arthritis, ulcerative colitis and Crohn's disease. Similar to LSMM, the cmFDR approach is a parametric method that integrates GWAS summary statistics and functional annotation information where functional annotation information provide `prior information' in a parametric two-group mixture model. The cmFDR approach assumes that compared to SNPs that are not functionally relevant, SNPs that are functionally relevant have a lower false discovery rate, and are associated with the trait. However, both cmFDR and LSMM do not exploit the pleiotropic relationship between traits with similar etiology to improve power to detect associations.

Finally, the third category of statistical methods combine the first two category criteria and integrate multiple GWAS trait data together with genotype-related functional annotation data. Two well known methods in this category include the genetic analysis incorporating pleiotropy and annotation (GPA) approach \cite{chung2014gpa} and the more recent latent probit model (LPM) approach \cite{Ming439133}. GPA employs a unified statistical framework to integrate genetically correlated GWAS traits by leveraging pleiotropy and functional annotation data to perform joint analysis of multiple traits. Similar to GPA, the three main goals of LPM are to identify the pleiotropic relationship between multiple traits by estimating the correlation between the traits, to identify the effect of functional annotations, and to improve the power to identify risk-associated SNPs for one or more traits. In both methods, the number of parameters that are included in the model increases significantly as the number of GWAS traits and functional annotations increase, rendering their implementation statistically and computationally challenging. Moreover, although methods in the second and third category can perform enrichment analysis on individual annotations, these methods do not consider interactions between the annotations, and therefore are limited in informing about the combined functional pathways through which genetic variants are associated with one or more traits. While some of these methods can theoretically be extended to include interactions between functional annotations to evaluate the combined functional effect of annotations, they retain the burden of knowing a priori the interactions that are of interest. Therefore, a method that can perform variable selection to identify relevant functional annotations or combinations of functional annotations from a large group of annotations that are linked to genetic variants associated with one or more traits is vitally important.

To address the statistical challenges and limitations described above, our team recently published a novel statistical approach called GPA-Tree \cite{gpatree} that simultaneously performs association mapping and identification of interactions between functional annotations. However, GPA-Tree does not exploit the pleiotropic relationship between two or more traits to improve association mapping power. In this work, we address the limitations of the GPA-Tree approach by proposing a new approach called multi-GPA-Tree. The multi-GPA-Tree approach is a novel statistical method based on a hierarchical modeling architecture, integrated with a multivariate regression tree algorithm \cite{de2002multivariate}. It exploits the pleiotropic relationship between traits with similar etiology to prioritize one or more trait-associated SNPs while simultaneously identifying key combinations of functional annotations related to the mechanisms through which one or more trait-associated SNPs influence the trait/s. Our comprehensive simulation studies and real data applications show that multi-GPA-Tree consistently improves statistical power to detect one or more trait-associated SNPs and also effectively identifies biologically important combinations of functional annotations. The multi-GPA-Tree approach takes GWAS summary statistics for multiple traits and functional annotation information for the GWAS genetic variants as input, and can be implemented using the R package \texttt{`multiGPATree'}. 

\section*{Materials and methods}
\subsection*{Overview of the multi-GPA-Tree approach}\label{multigpatreemodel}

% For figure citations, please use "Fig" instead of "Figure".
Let $\mathbf{Y}_{M \times D}$ be a matrix of genotype-trait association \textit{p}-values for $i = 1, 2,\cdots, M$ SNPs and $d = 1, 2,\cdots, D$ traits where $Y_{id}$ denotes the \textit{p}-value for the association of the $i^{th}$ SNP with the $d^{th}$ trait. 
$$
\mathbf{Y} = ({\mathbf{Y}}_{.1}, \ldots ,  {\mathbf{Y}}_{.D}) = \begin{pmatrix} 
y_{11} & \hdots & y_{1D} \\
\vdots & \ddots& \vdots \\
y_{M1} & \hdots & y_{MD} 
\end{pmatrix}_{M \times D}
$$
We also assume K binary annotations ($\mathbf{A}$) for each SNP.
$$
\mathbf{A} = ({\mathbf{A}}_{.1}, \ldots ,  {\mathbf{A}}_{.K}) = \begin{pmatrix} 
a_{11} & \hdots & a_{1K} \\
\vdots & \ddots& \vdots \\
a_{M1} & \hdots & a_{MK} 
\end{pmatrix}_{M \times K} \: \text{, \: where}
$$
 \begin{equation*}
  a_{ik} =
    \begin{cases}
      0, & \text{if $i^{th}$ SNP is not annotated in the $k^{th}$ annotation}\\
      1, & \text{if $i^{th}$ SNP is annotated in the $k^{th}$ annotation}
    \end{cases}       
\end{equation*}
To improve the power to identify risk-associated SNPs for one or more traits, GWAS association \textit{p}-values for $D$ traits ($\mathbf{Y}$) are integrated with functional annotations data ($\mathbf{A}$). The impact of functional annotations in modeling the relationship between GWAS traits and SNPs is characterized by defining a matrix  $\mathbf{Z}_{M \times 2^D} \in \{0, 1\}$  of latent binary variables where $\mathbf{Z}_i$ is a vector of length $2^D$ and indicates whether the $i^{th}$ SNP is null or non-null for the $D$ traits. Here, we present the model for the case of two GWAS traits ($D = 2$) to simplify notations. 

Let $Y \in \mathbf{R}^{M \times 2}$ be the matrix of GWAS association \textit{p}-values for two traits where $Y_{i1}$ and $Y_{i2}$ are the \textit{p}-values for the association between the $i^{th}$ SNP and traits $1$ and $2$, respectively. The latent binary vector is defined as $\mathbf{Z}_{i} = \{Z_{i00}, Z_{i10}, Z_{i01}, Z_{i11}\}$ for the $i^{th}$ SNP, where $Z_{i00} = 1$ indicates the $i^{th}$ SNP is null for both traits, $Z_{i10} = 1$ indicates the $i^{th}$ SNP is non-null for trait $1$ and null for trait $2$, $Z_{i01} = 1$ indicates the $i^{th}$ SNP is null for trait $1$ and non-null for trait $2$ and $Z_{i11} = 1$ indicates the $i^{th}$ SNP is non-null for both traits. We assume that a SNP can only be in one of the four states such that  $\sum\limits_{l \in \{00, 10, 01, 11\}} Z_{il} = 1$. The densities for SNPs in the null and non-null groups for both traits are assumed to come from $U[0, 1]$ and $Beta(\alpha_d, 1)$ distributions, where $0 < \alpha_d < 1$ and $d = 1, 2$, as proposed in Chung et al.\cite{chung2014gpa}. The distributions are defined as shown below.
\begin{equation*}
  \begin{split}
   & Y_{i1}|Z_{i00}=1 \sim U[0, 1] \hspace{1.8cm} Y_{i2}|Z_{i00}=0 \sim U[0, 1] \\
   & Y_{i1}|Z_{i10}=1 \sim Beta(\alpha_1, 1) \hspace{1.15cm} Y_{i2}|Z_{i10}=1 \sim U[0, 1] \\
   & Y_{i1}|Z_{i01}=1 \sim U[0, 1] \hspace{1.8cm} Y_{i2}|Z_{i01}=1 \sim Beta(\alpha_2, 1)\\
   & Y_{i1}|Z_{i11}=1 \sim Beta(\alpha_1, 1) \hspace{1.15cm} Y_{i2}|Z_{i11}=1 \sim Beta(\alpha_2, 1),
  \end{split}
\end{equation*} 
where $0 < \alpha_1, \alpha_2 < 1$. Finally, the functional annotation data $\mathbf{A}$ is integrated with the GWAS summary statistics data $\mathbf{Y}$ by defining a function $f$ that is a combination of functional annotations $\mathbf{A}$ and relating it to the multivariate expectation of latent $\mathbf{Z}$ as given in Eq \ref{eq1}.
\begin{equation}\label{eq1}
  \begin{split}
    P({Z}_{il}=1; a_{i1}, \ldots, a_{iK}) = \text{\it{f}}(a_{i1}, \ldots, a_{iK}), \text{where }l \in \{00, 10, 01, 11\}
  \end{split}
\end{equation}

For notational convenience we let $\boldsymbol{\theta}=(\alpha_1, \alpha_2)$ and denote $P({Z}_{il}=1; a_{i1}, \ldots, a_{iK})$ as $\boldsymbol{\pi_{.l}}$, where  $l \in \{00, 10, 01, 11\}$ such that $\boldsymbol{\pi}_{.00}$ are the prior probabilities that the SNPs are null for both traits, $\boldsymbol{\pi}_{.10}$ are the prior probabilities that the SNPs are non-null for trait $1$ and null for trait $2$, $\boldsymbol{\pi}_{.01}$ are the prior probabilities that the SNPs are null for trait $1$ and non-null for trait $2$, and $\boldsymbol{\pi}_{.11}$ are the prior probabilities that the SNPs are non-null for both traits. Then assuming that the SNPs are independent, the joint distribution of the observed data $Pr(\mathbf{Y}, \mathbf{A})$ and the incomplete and complete data log-likelihood can be written as shown in Eqs \ref{eq2}, \ref{eq3} and \ref{eq4}, respectively.
\begin{equation}\label{eq2} 
\begin{array}{l@{}l}
	Pr (\mathbf{Y}, \mathbf{A})
	&{}= \prod\limits_{i = 1}^{M} \bigg[ \sum\limits_{l \in \{00, 10, 01, 11\}}  P(Z_{il} = 1) P(Y_{i1}, Y_{i2}|Z_{il} = 1) \bigg] \\
	&{} =\prod\limits_{i = 1}^{M} \bigg[ \sum\limits_{l \in \{00, 10, 01, 11\} }  \pi_{il} \: P(Y_{i1}, Y_{i2}|Z_{il} = 1) \bigg]
	\end{array}
\end{equation}
\begin{equation}\label{eq3} 
\begin{array}{l@{}l}
	\ell_{IC} & {}= \sum\limits_{i = 1}^{M} log \bigg[ \sum\limits_{l \in \{00, 10, 01, 11\} }  \pi_{il} \: P(Y_{i1}, Y_{i2}|Z_{il} = 1) \bigg]
	\end{array}
\end{equation}
\begin{equation}\label{eq4}
\begin{array}{l@{}l}
\ell_{C}
&{}= \sum\limits_{i=1}^{M} \sum\limits_{l \in \{00, 10, 01, 11\} } Z_{il} \:  log\bigg[ \pi_{il} \: P(Y_{i1}, Y_{i2}|Z_{il} = 1) \bigg]
\end{array}
\end{equation}

\subsection*{Algorithm}\label{multigpatreealgorithm}

Given the approach described above, parameter estimation is implemented using an Expectation-Maximization (EM) algorithm \cite{moontk}. The function $f$ in Eq \ref{eq1} is estimated by using a multivariate regression tree algorithm\cite{de2002multivariate} that can identify combinations of functional annotations related to risk-associated SNPs for specific and multiple traits. The described approach is computationally implemented in two stages based on simulation study findings that showed improved parameter estimation and model stability when using a two-stage approach. Specifically, in Stage $1$, we first estimate the parameters $\alpha_1$ and $\alpha_2$ without identifying a combination of functional annotations. Then, in Stage $2$, we identify key combinations of functional annotations ($f(\mathbf{A})$) while the parameters $\alpha_1$ and $\alpha_2$ are kept fixed as the value obtained in Stage $1$. Detailed calculation steps are illustrated below.\\
\vspace{0.2cm}
\hspace{-0.5cm}\textbf{Stage 1:} In Stage 1, we initialize $\alpha_d^{(0)}=0.1$, $d = 1, 2$ and $\pi_{il}^{(0)} = \frac{1}{2^D}$, $D = 2$ (the number of traits). In the $t^{th}$ iteration of the E-step, define $Z_{il}^{(t)}, \: l \in \{00, 10, 01, 11\}$ for the $i^{th}$ SNP as:
\begin{equation}
\label{eq:aim2st1estep}
\begin{array}{l@{}l}
\hspace{-3.9cm}	\mathbf{E-step}: %\text{Define } Z_{il}^{(t)}, \: l \in \{00, 10, 01, 11\} \text{ as } \\
	z_{il}^{(t)} = P(Z_{il}=1 | \mathbf{Y}, \mathbf{A}; \: \boldsymbol{\theta}^{(t-1)}) \\

	\hspace{-1.5cm} = \frac{ \pi_{il}^{(t-1)} \: P(Y_{i1}, Y_{i2}|Z_{il} = 1;\: \boldsymbol{\theta}^{(t-1)} ) }{ \sum\limits_{l' \in \{00, 10, 01, 11\}}\pi_{il'}^{(t-1)} \: P(Y_{i1}, Y_{i2}|Z_{il'} = 1;\: \boldsymbol{\theta}^{(t-1)} ) } \\
\end{array}
\end{equation}
In the $t^{th}$ iteration of the M-step, ${\boldsymbol{\pi}_{i.}}$, $\alpha_1$ and $\alpha_2$ are updated as:
%shown in Equation  (\ref{eq:3}). \\
\begin{equation*}
\label{eq:3}
\begin{array}{l@{}l}
\bf{M-step}: \text{Fit a multivariate linear regression model as} \\
\hspace{3cm} \mathbf{Z}_{i.}^{(t)} = \beta_{0}^{(t)} + \beta_{1}^{(t)} {a}_{i1} + \cdots + \beta_K^{(t)} {a}_{iK} + {\epsilon_i}^{(t)} \\
\hspace{1.8cm} \text{Update } \boldsymbol{\pi}_{i.}  \text{as the predicted value from the multivariate linear} \\
\hspace{1.8cm}\text{regression model.} \\
%\hspace{2.2cm} \text{model.} \\
\hspace{1.8cm} \text{Update } \alpha_1^{(t)}= -\frac{\sum\limits_{i=1}^{M} ( z_{i10}^{(t)} \: + \: z_{i11}^{(t)} ) } {\sum\limits_{i=1}^{M} ( z_{i10}^{(t)} \: + \: z_{i11}^{(t)} )(logY_{i1}) } \text{ and } \alpha_2^{(t)}= -\frac{\sum\limits_{i=1}^{M} ( z_{i01}^{(t)} \: + \: z_{i11}^{(t)} ) } {\sum\limits_{i=1}^{M} ( z_{i01}^{(t)} \: + \: z_{i11}^{(t)} )(logY_{i2}) } 
\end{array}
\end{equation*} 
where $\beta_k^{(t)}, k = 0, \cdots, K$ are the regression coefficients and $\epsilon_i^{(t)}$ is the error term. 
%We have an explicit solution for the $\alpha$ update, as derived by maximizing the complete data log-likelihood shown in Section 1 in the Supplementary Materials. 
The E and M steps are repeated until the incomplete log-likelihood and the $\alpha_1$ and $\alpha_2$ estimates converge. Then, $\alpha_1$, $\alpha_2$ and $\boldsymbol{\pi}_{i.}$ estimated in this stage are used to fix $\alpha_1$, $\alpha_2$ and initialize $\boldsymbol{\pi}_{i.}$, respectively, in Stage $2$.\\
\vspace{0.2cm}
\hspace{-0.5cm}\textbf{Stage 2:} In stage 2, we implement another EM algorithm employing the multivariate regression tree algorithm, which allows for identification of union, intersection, and complement relationships between functional annotations in estimating $\boldsymbol{\pi}_{i.}$. In the $t^{th}$ iteration of the E-step, define $Z_{il}^{(t)}, \: l \in \{00, 10, 01, 11\}$ for the $i^{th}$ SNP as shown in Eq \ref{eq:aim2st1estep}, except $\alpha_1$ and $\alpha_2$ are fixed as $\hat{\alpha_1}$ and $\hat{\alpha_2}$, which are the final estimates of $\alpha_1$ and $\alpha_2$ obtained from Stage $1$. 

\begin{equation*}
\label{eq:48}
\begin{array}{l@{}l}
\mathbf{E-step}: \text{Define } Z_{il}^{(t)}, \: l \in \{00, 10, 01, 11\} \text{ as in Eq \ref{eq:aim2st1estep}, except $\alpha_1$ and $\alpha_2$ are } \\
\hspace{1.8cm} \text{fixed as $\hat{\alpha_1}$ and $\hat{\alpha_2}$, the final estimates of $\alpha_1$ and $\alpha_2$ from Stage 1.} 
\end{array}
\end{equation*}
In the $t^{th}$ iteration of the M-step, $\boldsymbol{\pi}_{i.}$ is updated as:
\begin{equation}\label{eq6}
\begin{array}{l@{}l}
\mathbf{M-step}: \text{Fit a multivariate regression tree model as shown below.}\\ 
\hspace{2.8cm} \mathbf{Z}_{i.}^{(t)} = {f}^{(t)}(a_{i1}, \cdots, a_{iK}) + \epsilon_i^{(t)}, \text{ where } \epsilon_i \text{ is the error term}. \\
\hspace{1.8cm} \text{Update } {\boldsymbol{\pi}_{i.}}^{(t)} \text{ as the predicted values from the multivariate regression} \\
\hspace{1.8cm} \text{tree model.}
\end{array}
\end{equation}
 In the M-step, the complexity parameter ($cp$) of the multivariate regression tree is the key tuning parameter and defined as the minimum improvement that is required at each node of the tree. Specifically, in the multivariate regression tree model, the largest possible tree (i.e., a full-sized tree) is first constructed and then pruned using $cp$. This approach allows for the construction of the accurate yet interpretable multivariate regression tree that can explain relationships between functional annotations and risk-associated SNPs for one or more traits.  The E and M steps are repeated until the incomplete log-likelihood converges. The pruned tree structure identified by the multivariate regression tree model upon convergence of the Stage 2 EM is the $f$ in Eq \ref{eq1}. 

We note that unlike the standard EM algorithm, the incomplete log-likelihood in Stage $2$ is not guaranteed to be monotonically increasing. Therefore, we implement Stage $2$ as a generalized EM algorithm by retaining only the iterations in which the incomplete log-likelihood increases compared to the previous iteration.

\subsection*{Prioritization of marginal and joint risk-associated SNPs and identification of relevant functional annotations}
Following parameter estimation, we can prioritize one or more trait risk associated SNPs using local false discovery rate or $fdr$. As shown in Eq \ref{eq7}, for marginal associations with a specific trait, we define $fdr$ as the marginal posterior probability that the $i^{th}$ SNP belongs to the non-risk-associated group for the specific trait given its GWAS association \textit{p}-values for all traits and functional annotation information. Likewise, for joint associations between traits, we define $fdr$ as the joint posterior probability that the $i^{th}$ SNP belongs to the non-risk-associated group for the traits given its GWAS association \textit{p}-values for all traits and functional annotation information. Next, we utilize the `direct posterior probability' approach \cite{newton2004detecting} to control the global false discovery rate (FDR).
\begin{equation}\label{eq7}
\begin{array}{l@{}l}
fdr_1(\mathbf{Y}_{i.}, \mathbf{A}_{i.}) = P(Z_{i00} + Z_{i01} = 1 | \mathbf{Y}_{i.}, \mathbf{A}_{i.}, \: \hat{\boldsymbol{\theta}} ) = \frac{P(Y_{i1}, Y_{i2}, Z_{i00} + Z_{i01} = 1; \: \hat{\boldsymbol{\theta} })}{P(Y_{i1}, Y_{i2}; \: \hat{\boldsymbol{\theta}} ) }, \\
fdr_2(\mathbf{Y}_{i.}, \mathbf{A}_{i.}) = P(Z_{i00} + Z_{i10} = 1 | \mathbf{Y}_{i.}, \mathbf{A}_{i.}, \: \hat{\boldsymbol{\theta}} ) = \frac{P(Y_{i1}, Y_{i2}, Z_{i00} + Z_{i10} = 1; \: \hat{\boldsymbol{\theta} })}{P(Y_{i1}, Y_{i2}; \: \hat{\boldsymbol{\theta}} ) }, \\
fdr_{1,2}(\mathbf{Y}_{i.}, \mathbf{A}_{i.}) = P(Z_{i00} + Z_{i10}+ Z_{i01} = 1| \mathbf{Y}_{i.}, \mathbf{A}_{i.} ) = \frac{P(Y_{i1}, Y_{i2}, Z_{i00} + Z_{i10}+ Z_{i01} = 1; \: \hat{\boldsymbol{\theta} })}{P(Y_{i1}, Y_{i2}; \: \hat{\boldsymbol{\theta}} ) }, \\
\text{where} \\

\hspace{2.7cm} P(Y_{i1}, Y_{i2}; \: \hat{\boldsymbol{\theta}} )  = \sum\limits_{l \in \{00, 10, 01, 11 \}} \hat{\pi}_{il}\: P(Y_{i1}, Y_{i2} | Z_{il}, \mathbf{A}_{i.}; \: \hat{\boldsymbol{\theta}} ) ,\\

P(Y_{i1}, Y_{i2}, Z_{i00} + Z_{i01} = 1 ; \: \hat{\boldsymbol{\theta}} )  = \sum\limits_{l \in \{00, 01 \}} \hat{\pi}_{il}\: P(Y_{i1}, Y_{i2} | Z_{il}, \mathbf{A}_{i.}; \: \hat{\boldsymbol{\theta}} ) ,\\

P(Y_{i1}, Y_{i2}, Z_{i00} + Z_{i10} = 1; \: \hat{\boldsymbol{\theta}} )  = \sum\limits_{l \in \{00, 10\}} \hat{\pi}_{il}\: P(Y_{i1}, Y_{i2} | Z_{il}, \mathbf{A}_{i.}; \: \hat{\boldsymbol{\theta}} ), \\

P(Y_{i1}, Y_{i2}, Z_{i00} + Z_{i10} + Z_{i01} = 1; \: \hat{\boldsymbol{\theta}} )  = \sum\limits_{l \in \{00, 10, 01\}} \hat{\pi}_{il}\: P(Y_{i1}, Y_{i2} | Z_{il}, \mathbf{A}_{i.}; \: \hat{\boldsymbol{\theta}} ), \\
\end{array}
\end{equation}

 Finally, relevant combinations of functional annotations are inferred based on the combination of functional annotations selected by the multivariate regression tree model upon convergence of the Stage 2 EM algorithm.  

% Results and Discussion can be combined.

\section*{Results}

\subsection*{Simulation study}
\begin{figure}[!h]
\centerline{\includegraphics[width=0.75\textwidth,height=0.75\textheight,keepaspectratio]{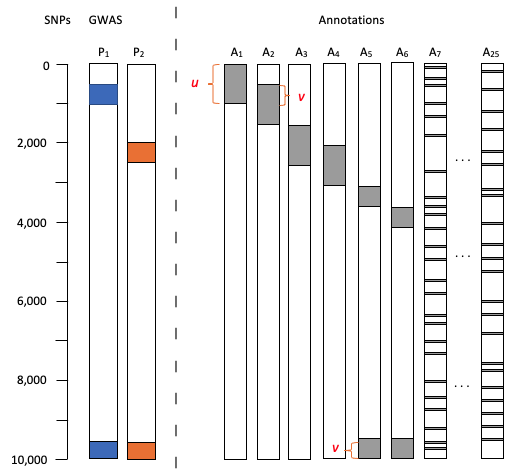}}
\caption{{\bf Simulation setting.}
The graphical scenario is presented for $M$ = $10,000$ SNPs; $K$ = $25$ annotations; $\%$ of annotated SNPs in $A_1$-$A_6$ ($u$) = $10\%$; $\%$ of overlap between $A_1$-$A_2$, $A_3$-$A_4$, $A_5$-$A_6$ ($v$)= $50\%$; $A_7$–$A_{15}$ are noise SNPs, approximately $20\%$ of which are randomly annotated; blue SNPs are non-null for trait $P_1$ and their GWAS p-values are generated from $ Beta(\alpha_1=0.4, 1)$ distribution; orange SNPs are non-null for trait $P_2$ and their GWAS p-values are generated from $ Beta(\alpha_2=0.4, 1)$ distribution; all other SNPs are null for both traits and their GWAS p-values are generated from $U[0,1]$ distribution for both traits.}
\label{fig1}
\end{figure}

We conducted a simulation study to evaluate the performance of the proposed multi-GPA-Tree approach. Fig~\ref{fig1} provides a graphical depiction of the simulation setting. For all simulation data, the number of SNPs was set to $M = 10,000$, the number of annotations was set to $K = 25$, SNPs that are marginally associated with the first trait ($P_1$) were assumed to be characterized with the combinations of functional annotations defined by $L_1 = A_1 \cap A_2$, SNPs that are marginally associated with the second trait ($P_2$) were assumed to be characterized with the combinations of functional annotations defined by $L_2 = A_3 \cap A_4$, SNPs that are jointly associated with traits $P_1$ and $P_2$ were assumed to be characterized with the combinations of functional annotations defined by $L_3 = A_5 \cap A_6$,  all the remaining functional annotations ($A_k, k = 7, \ldots, 25$) were considered to be noise annotations. Approximately $10\%$ of SNPs were assumed to be annotated for annotations $A_1 - A_6$, and  $v\% $ where $v = 35\%, 50\%$ and $75\%$ of those annotated were assumed to overlap between the true combinations of functional annotations. For noise annotations $A_7 - A_{25}$, approximately $20\%$ of SNPs were annotated by first generating the proportion of annotated SNPs from $Unif[0.1, 0.3]$ and then randomly setting this proportion of SNPs to one. For trait $P_1$, the SNPs that satisfied the functional annotation combination in $L_1$ or $L_3$ were assumed to be risk-associated SNPs and their \textit{p}-values were simulated from $Beta(\alpha_1, \: 1)$ with $\alpha_1 = 0.4$. Similarly, for trait $P_2$, the SNPs that satisfied the functional annotation combination in $L_2$ or $L_3$ were assumed to be risk-associated SNPs and their \textit{p}-values were simulated from $Beta(\alpha_2, \: 1)$ with $\alpha_2 = 0.3$. The SNPs that did not satisfy the required condition for association with $P_1$ or $P_2$ were assumed to be non-risk SNPs and their \textit{p}-values were simulated from $U[0, \: 1]$. 

\begin{figure}[!h]
\centerline{\includegraphics[width=1\textwidth,height=1\textheight,keepaspectratio]{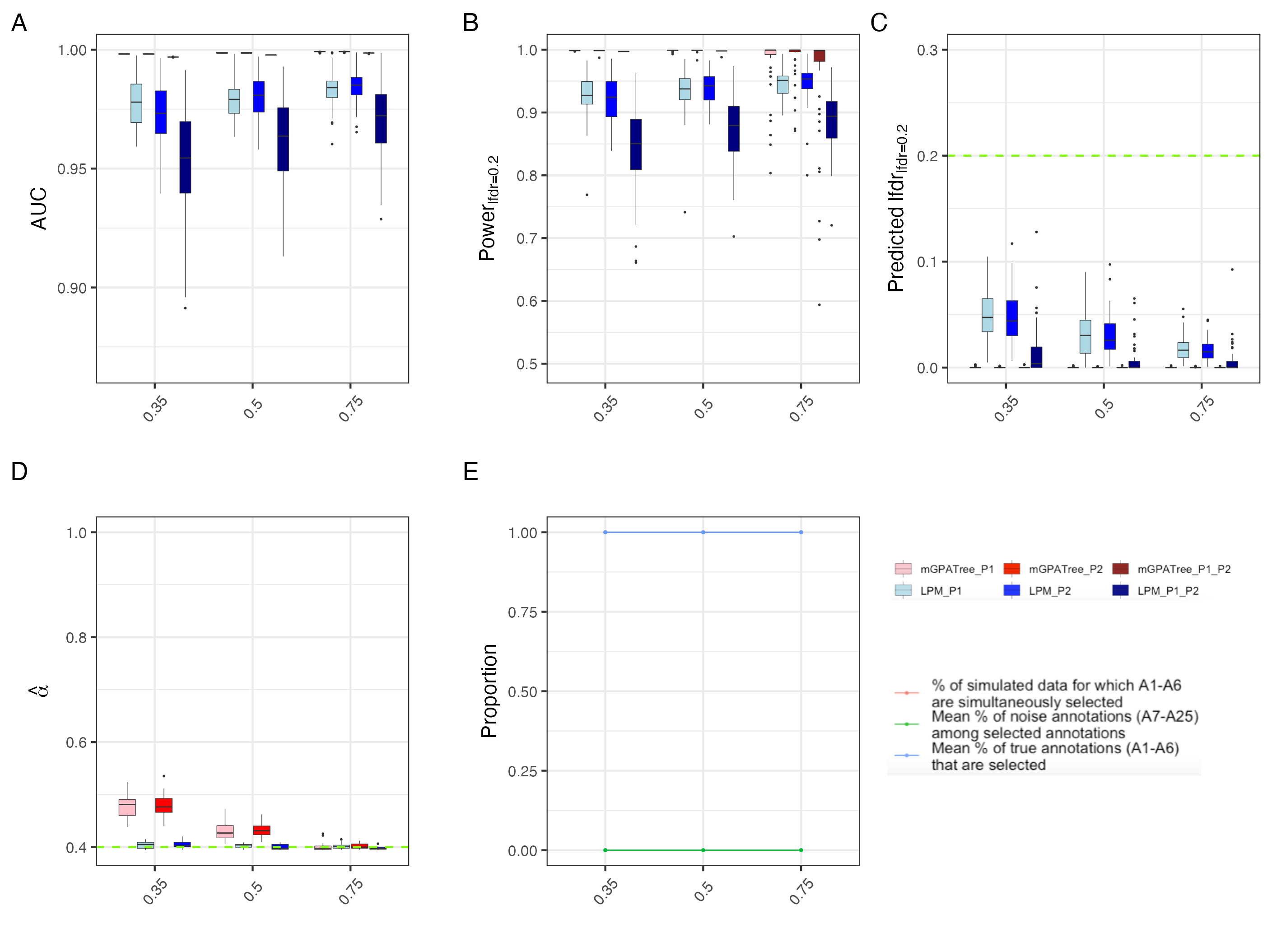}}
\caption{{\bf Simulation results.}
Comparison of (A) AUC, (B) statistical power to detect true marginal and joint risk-associated SNPs when local FDR ($lfdr$) is controlled at the nominal level of $0.20$, (C) predicted local FDR ($lfdr$) when controlled at the nominal level of $0.20$, and (D) estimated $\alpha_1$ and $\alpha_2$ parameter for traits P1 and P2 between multi-GPA-Tree and LPM; and (E) proportion of simulated data for which only true annotations ($A_1-A_6$) are simultaneously selected (red line),  the average proportion of noise annotations ($A_7-A_{25}$) among the functional annotations identified by multi-GPA-Tree (green line), and the average proportion of true  annotations $A_1-A_6$ among the annotations identified by multi-GPA-Tree (blue line). We note that the blue and red lines are overlaid in the plot. The results are presented for different proportions of the overlap between SNPs annotated in $A_1 - A_2$, $A_3 - A_4$ and $A_5 - A_6$ ($v$; x-axis). $M = 10,000$, $K = 25$, and $\alpha_d =0.4$ in $Beta(\alpha_d, 1), d=1,2$. Results are summarized from 50 replications. Results related to marginal associations are presented using suffix *$\_$P1 and *$\_$P2 and results related to joint associations are presented using suffix *$\_$P1$\_$P2.}
\label{fig2}
\end{figure}

We simulated $50$ datasets and compared the performance of multi-GPA-Tree with LPM \cite{Ming439133} using the simulation parameters defined above.  The metrics for comparing the methods included (1) area under the curve (AUC) for marginal and joint associations, where the curve was created by plotting the true positive rate (sensitivity) against the false positive rate (1-specificity) to detect one or more trait risk-associated SNPs when global FDR was controlled at various levels; (2) statistical power to identify marginal and joint risk-associated SNPs when local FDR ($lfdr$) was controlled at the nominal level of 0.20; (3) predicted $lfdr$ when $lfdr$ was controlled at the nominal level of $0.20$; and (4) estimation accuracy for $\alpha_d$ parameters in the $Beta(\alpha_d, \: 1) , d = 1, 2$ distribution used to generate the \textit{p}-values of risk-associated groups for traits $P_1$ and $P_2$.  For multi-GPA-Tree, we also examined the accuracy of detecting the correct functional annotation tree based on (1) the proportion of simulation data for which all relevant functional annotations in $L_1$, $L_2$ and $L_3$, i.e, annotation $A_1-A_6$, were identified simultaneously; (2) the average proportion of noise functional annotations ($A_7-A_{25}$) among the functional annotations identified by multi-GPA-Tree; and (3) the average proportion of true functional annotations ($A_1-A_6$) among the functional annotations identified by multi-GPA-Tree. Here we especially investigated how the the overlap between SNPs annotated in $A_1 - A_2$, $A_3 - A_4$ and $A_5 - A_6$ ($v$) impacted multi-GPA-Tree's ability to separate relevant functional annotations from noise annotations for one or more trait risk-associated SNPs.

\begin{itemize}
    \item {\bf AUC: } Fig \ref{fig2}A compares the distribution of AUCs returned by multi-GPA-Tree and LPM. For all $v$, multi-GPA-Tree showed consistently higher AUC relative to LPM for both marginal and joint association. LPM showed higher AUC for marginal associations relative to joint association.
    
    \item {\bf Statistical power: } Fig \ref{fig2}B compares the distribution of power to detect true marginal and joint risk-associated SNPs when local FDR ($lfdr$) was controlled at 0.20 between multi-GPA-Tree and LPM. The multi-GPA-Tree approach showed higher statistical power to detect true marginal and joint risk-associated SNPs relative to LPM for all $v$. LPM showed higher power for marginal associations relative to joint association. LPM showed greater variability in statistical power compared to multi-GPA-Tree overall while multi-GPA-Tree showed more variability in power for higher $v$.
    
    \item {\bf Predicted local fdr ($\mathbf{lfdr}$)}: Fig \ref{fig2}C compares the distribution of predicted $lfdr$ between multi-GPA-Tree and LPM when $lfdr$ was controlled at the nominal level of $0.20$. Although LPM showed higher perdicted $lfdr$ compared to multi-GPA-Tree, both multi-GPA-Tree and LPM showed consistently controlled $lfdr$ under $0.20$ at the $0.20$ level for all $v$.
    
    \item {\bf Estimation of $\boldsymbol{\alpha}$ parameters:} Fig \ref{fig2}D shows the distribution of $\alpha$ parameter estimates for traits 1 and 2 (P1 and P2) using multi-GPA-Tree and LPM. LPM was on average more accurate than multi-GPA-Tree in estimating $\alpha$ for both traits. The multi-GPA-Tree approach generally overestimated $\alpha$ and this was most notable for smaller $v$. As $v$ increased, $\alpha$ estimates from multi-GPA-Tree became closer to the true value. We note that overestimation of $\alpha$ by multi-GPA-Tree did not impact the method’s ability to identify the true combinations of functional annotations or the marginal and joint risk-associated SNPs, which are the main objectives of multi-GPA-Tree.
    
    \item {\bf Selection of relevant and noise annotations:} The red line in Fig \ref{fig2}E shows the proportion of times only functional annotations in the true combination $L_1$, $L_2$ and $L_3$ ($A_1 - A_6$) were simultaneously identified by multi-GPA-Tree. The red line aligned exactly with the blue line which shows the mean proportion of true annotations ($A_1 - A_6$) among all selected annotations. Finally, the green line shows the proportion of noise annotations ($A_7 - A_{25}$) among the selected annotation. The alignment of the red and blue lines and the horizontal green line at $0$ suggest that only and all relevant annotations were selected by multi-GPA-Tree. These results demonstrate the potential of multi-GPA-Tree to correctly identify true annotations from noise annotations.

\end{itemize}

\subsection*{Real data application}

\begin{figure}[!h]
\includegraphics[width=1\textwidth,height=1\textheight,keepaspectratio]{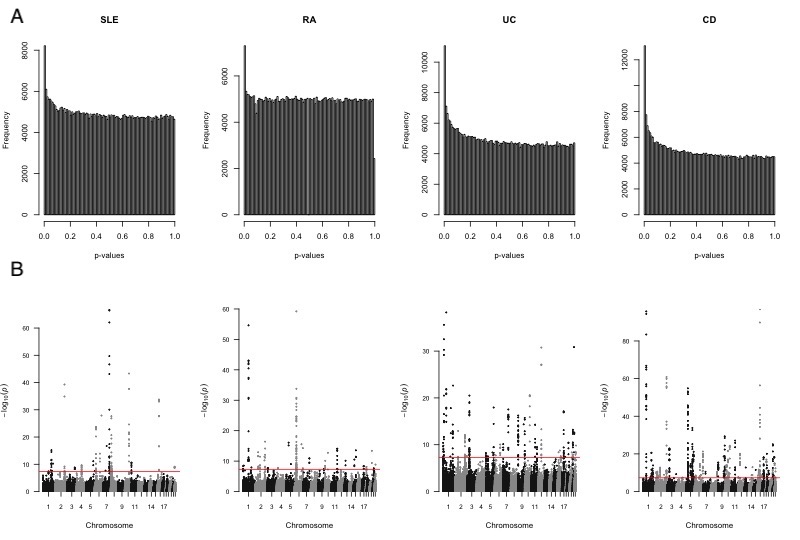}
\caption{{\bf GWAS summary statistic plots.}
(A) GWAS p-value histogram and (B) Manhattan plots for the four GWAS. Genome-wide significance level ($-log_{10}(5 \times 10^{-8})$) is indicated by the red line.}
\label{fig:hist_manhattan_plot}
\end{figure}

We obtained a combined dataset including the SLE \cite{langefeld2017transancestral} and RA \cite{okada2014genetics}, and CD and UC \cite{de2017genome} GWAS. Summary statistics in the SLE and RA GWAS was profiled for $18,264$ ($6,748$ cases and $11,516$ controls) and $58,284$ ($14,361$ cases and $43,923$ controls) individuals of European ancestry, respectively. Summary statistics in the CD and UC GWAS was profiled from $8,467$ ($4,686$ cases and $3,781$ controls) individuals of European ancestry. Following quality control and exclusion of SNPs in the MHC region, approximately $492,557$ SNPs were utilized in the final analysis and integrated with functional annotation data from GenoSkyline (GS) \cite{lu2016integrative} and GenoSkylinePlus (GSP) \cite{lu2017systematic}. The Manhattan plots and \textit{p}-value histogram plots for the four GWAS data are presented in Fig \ref{fig:hist_manhattan_plot}A and \ref{fig:hist_manhattan_plot}B, respectively.

\begin{figure}[!ht]
%\centerline{\includegraphics{fig01.eps}}
\includegraphics[width=1\textwidth,height=1\textheight,keepaspectratio]{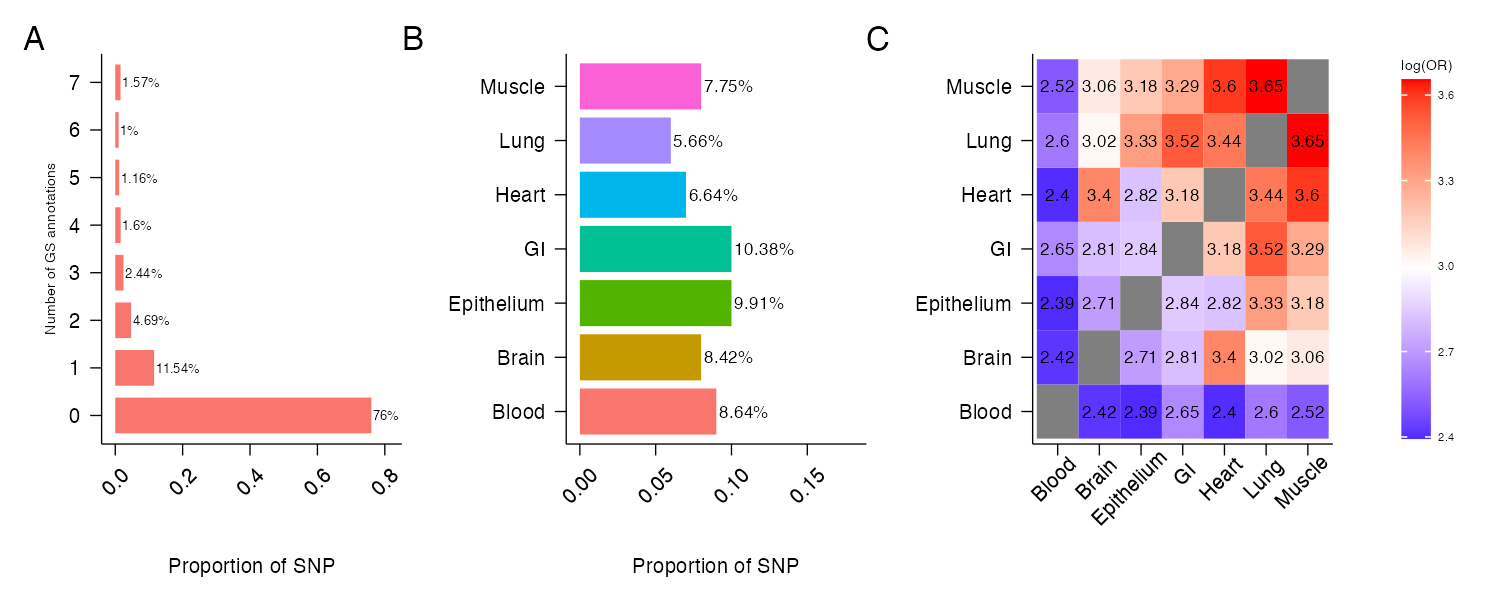}
\caption{Characteristics of $492,557$ SNPs when integrated with seven GenoSkyline (GS) annotations. (A) Number of GS tissues in which SNPs are annotated. (B) Proportion of SNPs that are annotated for each GS tissue type. (C) Overlap of SNPs annotated by seven GS tissue types, calculated using log odds ratio.}
\label{fig:aim2rdaAllGS}
\end{figure}

\begin{figure}[!ht]
%\centerline{\includegraphics{fig01.eps}}
\includegraphics[width=1\textwidth,height=1\textheight,keepaspectratio]{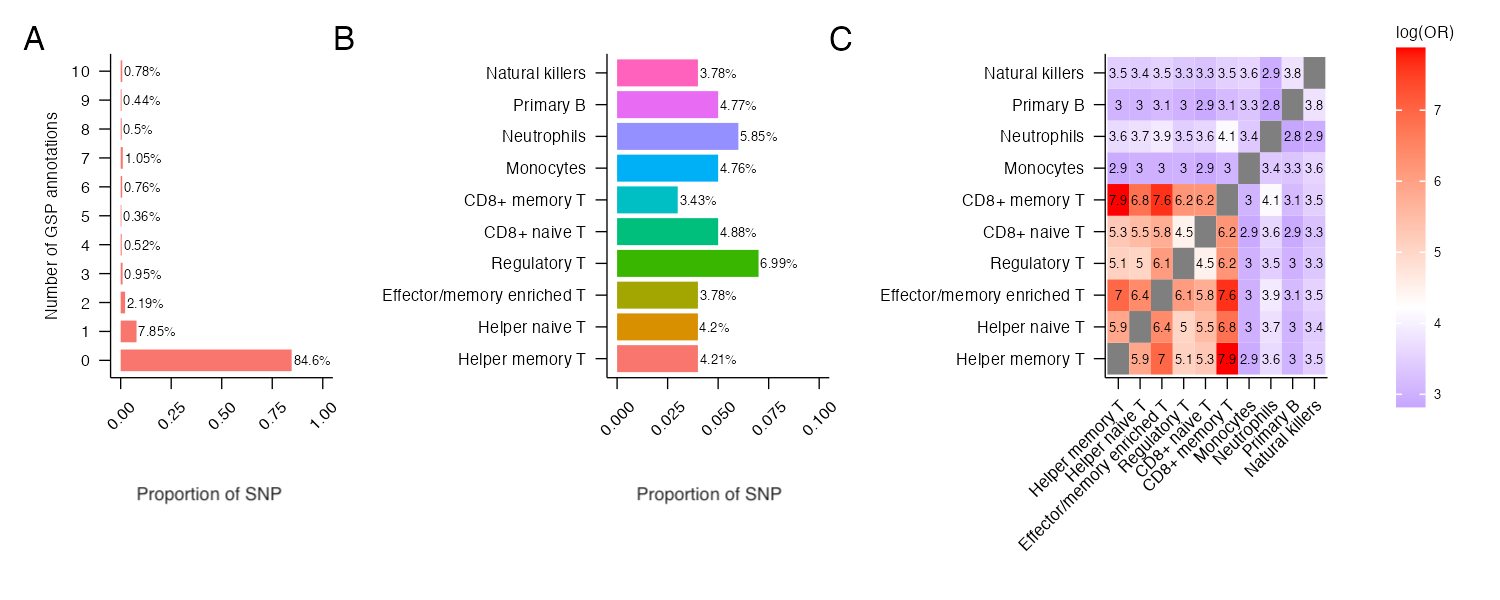}
\caption{Characteristics of $492,557$ SNPs when integrated with 10 blood related GenoSkylinePlus (GSP) annotations. (A) Number of GSP tissues in which SNPs are annotated. (B) Proportion of SNPs that are annotated for each blood related GSP annotations. (C) Overlap of SNPs annotated by 10 blood related GPS annotations, calculated using log odds ratio.}
\label{fig:aim2rdaAllGSP}
\end{figure}

We descriptively investigated the functional potential of the $492,557$ SNPs using seven tissue-specific GS annotations (Fig \ref{fig:aim2rdaAllGS}) and ten blood-related cell-type specific GSP annotations (Fig \ref{fig:aim2rdaAllGSP}). With a GS and GSP score cutoff of $0.5$, $24\%$ of SNPs were annotated in at least one of the seven tissue types (Fig \ref{fig:aim2rdaAllGS}A) and $15.4\%$ of SNPs were annotated in at least one of the 10 blood related cell-type specific annotations (Fig \ref{fig:aim2rdaAllGSP}A). The percentage of annotated SNPs ranged from $5.66\%$ for lung tissue to $10.38\%$ for GI tissue (Fig \ref{fig:aim2rdaAllGS}B) and from $3.43\%$ for primary T CD8$^+$ memory cells to $6.99\%$ for primary T regulatory cells (Figure \ref{fig:aim2rdaAllGSP}B). We also measured the overlap in SNPs annotated in different tissue-types and cell-types using log odds ratio (Fig \ref{fig:aim2rdaAllGS}C and \ref{fig:aim2rdaAllGSP}C). Consistent with the literature stipulating that muscle and lung tissues show higher levels of eQTL sharing while blood shows the lowest \cite{gtex2015genotype, lu2016integrative}, our findings show that SNPs annotated for muscle, lung and heart tissues overlap more with other tissue types while SNPs annotated for blood tissue overlap less (Fig \ref{fig:aim2rdaAllGS}C). Finally, we observed the different types of T cells (Primary helper memory, helper naive, effector/memory enriched, regulatory, CD8$^+$ naive and CD8$^+$ memory T cells) overlap more with each other while neutrophils, primary B and natural killer cells overlap less (Fig \ref{fig:aim2rdaAllGSP}C). 

\subsubsection*{Integration of Systemic Lupus Erythematosus (SLE) and Rheumatoid Arthritis (RA) GWAS}

\subsubsection*{\textit{Tissue-level investigation using GenoSkyline (GS) annotations}}

We applied the multi-GPA-Tree approach to the SLE and RA GWAS and tissue-specific GS annotations to identify SNPs that are marginally and jointly associated with SLE and RA, and to characterize the functional annotations relevant to single and multiple trait risk-associated SNPs. At the nominal global FDR level of $0.05$, multi-GPA-Tree identified $394$ SNPs that are jointly associated with both SLE and RA, $1087$ SNPs that are marginally associated with SLE and $791$ SNPs that are marginally associated with RA (Table \ref{reftab1}).

\begin{figure}[!h]
\includegraphics[width=1\textwidth,height=1\textheight,keepaspectratio]{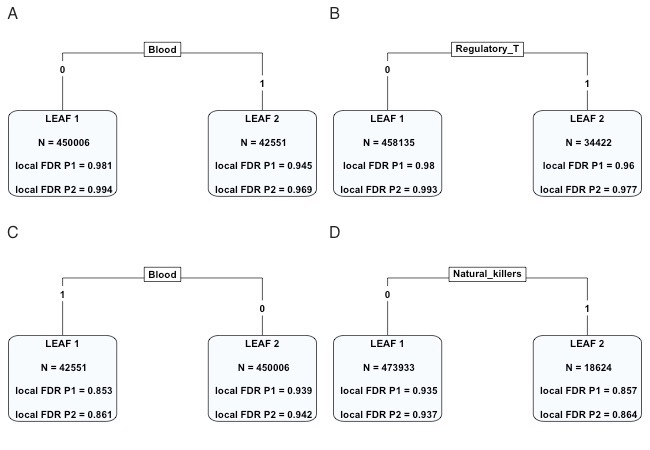}
\caption{{\bf Real data application results.} Trees returned by multi-GPA-Tree models when (A) SLE (P1), RA (P2) and GenoSkyline annotations are integrated, (B) SLE (P1), RA (P2) and GenoSkylinePlus annotations are integrated, (C) UC (P1), CD (P2) and GenoSkyline annotations are integrated, and (D) UC (P1), CD (P2) and GenoSkylinePlus annotations are integrated.}
\label{fig:tree_plot}
\end{figure}

\begin{table}[!ht]
\resizebox{\textwidth}{!}{%
\begin{tabular}{ |p{2.2cm}||p{2.5cm}|p{2cm}|p{2cm}|p{2.5cm}|p{3cm}|  }
  %\multicolumn{5}{|c|}{Country List} \\
 \hline
 Data integration & Approach & $\#$ marginally associated with P1 (SLE/UC) & $\#$ marginally associated with P2 (RA/CD) & $\#$ jointly associated with P1 and P2 (SLE+RA/ UC+CD) & Selected annotation\\
 \hline
 \hline
SLE+RA+GS   &  multi-GPA-Tree   & 1087&  791& 394 & Blood\\
%\hline
SLE+GS & GPA-Tree & 696 & - & - & Blood\\
%\hline
RA+GS & GPA-Tree &- & 470 & - & Blood\\
\hline
%\hline
SLE+RA+GSP & multi-GPA-Tree   & 1065 & 760 &  383 & Regulatory T\\
%\hline
SLE+GSP & GPA-Tree    & 830 & - & - & Primary B\newline Regulatory T\newline Helper memory T\\
%\hline
RA+GSP & GPA-Tree & - & 634& - & Regulatory T\newline Helper memory T\newline Natural killer\\
\hline
%\hline
UC+CD+GS & multi-GPA-Tree  & 5430&  5041& 5041 & Blood\\
UC+GS & GPA-Tree & 1566 & - & - & Blood\\
CD+GS & GPA-Tree & - & 3185 & - & Blood\\
\hline
UC+CD+GSP & multi-GPA-Tree & 4995 &  4912 & 4576 & Natural killer\\
UC+GSP & GPA-Tree & 1654 & - & - & Natural killer \newline Monocytes \newline Effector/Memory enriched T \newline Regulatory T\\
CD+GSP & GPA-Tree & - & 3232 & - & Natural killer \newline Monocytes \newline Effector/Memory enriched T \newline Primary B \newline Regulatory T \\
 \hline
\end{tabular}
}
\caption{{\bf Real data application results summary.}
Number of jointly and marginally associated SNPs when systemic lupus erythematosus (SLE) and rheumatoid arthritis (RA), and Crohn's disease (CD) and ulcertive colitis (UC) GWAS are integrated with the GenoSkyline (GS) and GenoSkylinePlus (GSP) annotations when jointly and individually analyzed using the multi-GPA-Tree and the GPA-Tree approach. All analysis included $492,557$ SNPs and $7$ tissue-specific GS and $10$ blood-related cell-type specific GSP annotations. Inference is based on global FDR control at the nominal level of 0.05.}
\label{reftab1}
\end{table}

In the joint analysis of SLE and RA with tissue-specific GS annotations, the original multi-GPA-Tree model identified blood tissue at the root node and included $2$ leaves (Fig. \ref{fig:tree_plot}A). Further investigation showed that $156$ SNPs that were jointly associated with both SLE and RA, $336$ SNPs that were marginally associated with SLE and $306$ SNPs that were marginally associated with RA were also annotated for blood tissue. Of the $156$ jointly associated SNPs that were also annotated for blood tissue, $118$ SNPs were protein-coding such that chromosomes $1, 6, 2$ and $17$ had the most number of protein-coding SNPs. The \textit{PLCL1} gene in chromosome $2$, \textit{IL2RA} gene in chromosome $10$ and \textit{UHRF1BP1} gene in chromosome $6$ had the most number of protein-coding SNPs with $5$ SNPs related to coding the \textit{PLCL1} gene and $4$ SNPs each related to coding the \textit{IL2RA} and \textit{UHRF1BP1} genes. The \textit{PLCL1} gene is known to promote inflammatory response by regulating the NLRP3 inflammasomes, a component of the immune system related to activation and secretion of proinflammatory cytokines \cite{luos}. Similarly, \textit{IL2RA} gene expression has been reported on activated T and B cells, regulatory T cells, activated monocytes, and natural killer cells \cite{carrej, caruso}, and the \textit{UHRF1BP1} gene plays a role in non-conservative amino-acid change and is related to RNA processing complex that is targeted by SLE autoantibodies \cite{gateva}.

We also discovered $3$ SNPs each in chromosomes $4, 17, 3, 3,$ and $16$ known to code the \textit{BANK1}, \textit{PGAP3}, \textit{PLCL2}, \textit{RASA2}, and \textit{TXNDC11} genes, respectively. \textit{BANK1} is primarily expressed in CD19$^+$ B cells and is a known SLE and RA susceptibility gene \cite{kozyrev, orozco}; in animal studies, the \textit{PGAP3} gene knockout has been associated with reduced apoptotic cell clearance, a causal pathway for autoimmunity \cite{wangy}; \textit{PLCL2} is known to encode a negative regulator of B cell receptor signalling important in controlling immune responses and is a known susceptibility gene for RA \cite{bowesj}. Finally, although not explored in the context of SLE and RA, \textit{RASA2} variants are known to be associated with combined allergy diseases \cite{ferreira} and \textit{TXNDC11} is known to play a role in thyroid hormone biosynthesis \cite{jaeger}. \\

We additionally implemented the GPA-Tree approach by integrating the GS annotations to the SLE and RA GWAS individually. Validating our multi-GPA-Tree results, blood tissue was identified at the root node in the separate GPA-Tree analysis for both SLE and RA. In the individual GPA-Tree analysis, we identified $696$ SNPs to be associated with SLE and $470$ SNPs to be associated with RA with $229$ SLE associated and $224$ RA associated SNPs also annotated for blood tissue. Of the top $3$ genes identified in the joint analysis of SLE and RA, one or more protein-coding SNPs related to the \textit{IL2RA} and \textit{PLCL1} genes were also identified in the single trait analysis of both SLE and RA. However, protein-coding SNPs related \textit{UHRF1BP1} gene were identified for SLE but not for RA in single trait analysis.

\subsubsection*{\textit{Cell-type-level investigation using GenoSkylinePlus (GSP) annotations}}

Based on the observed relationship between GS annotation for blood tissue and SLE and RA, in the second phase of the analysis, we applied the multi-GPA-Tree approach to the SLE and RA GWAS and 10 blood related cell-type specific GSP annotations to identify SNPs that were marginally and jointly associated with SLE and RA, and to characterize the blood related GSP functional annotations relevant to single and multiple trait risk-associated SNPs. At the nominal global FDR level of $0.05$, multi-GPA-Tree identified $383$ SNPs that were jointly associated with SLE and RA,  $1,065$ SNPs that were marginally associated with SLE and $760$ SNPs that were marginally associated with RA (Table \ref{reftab1}). The joint analysis also identified primary T regulatory cells at the root node (Fig. \ref{fig:tree_plot}B) with $95$ SNPs that were jointly associated with both SLE and RA, $191$ SNPs that were marginally associated with SLE and $176$ SNPs that were marginally associated with RA also annotated for regulatory T cells. Of the $95$ jointly associated SNPs that were also annotated for regulatory T cells, $69$ were protein coding such that chromosomes $1, 6, 16$ and $2$ had the most number of protein-coding SNPs. The \textit{PLCL1} gene in chromosome $2$, \textit{IL2RA} gene in chromosome $10$ and \textit{TXNDC11} gene in chromosome $16$ had the most number of protein-coding SNPs with $3$ different protein-coding SNPs related to coding each of the $3$ genes.\\

The individual analysis using the GPA-Tree approach identified primary B, regulatory T and helper memory T cells with primary B cell at the root node for SLE. Similarly, we identified regulatory T, helper memory T and natural killer cells with regulatory T cells at the root node for RA. In the individual GPA-Tree analysis, we identified $830$ SNPs to be associated with SLE and $634$ SNPs to be associated with RA. Of those associated with SLE, 176 were annotated for primary B, 122 were annotated for regulatory T, and 43 were annotated for helper memory T cells. Among SNPs associated with RA, 132 were annotated for both regulatory T and natural killer cells, 148 were annotated for regulatory T and not for natural killer cells, 32 were annotated for natural killer but not for regulatory T cells and 35 were annotated for helper memory T cells. Among the top 3 genes identified in the joint analysis of SLE, RA and GSP annotations, one or more protein-coding SNPs related to the \textit{IL2RA} gene were also identified in the single trait analysis of both SLE and RA. However, protein-coding SNPs related to the \textit{PLCL1} gene were identified for SLE only while protein-coding SNPs related to the \textit{TXNDC11} gene were identified for RA only. 

\subsubsection*{Integration of Ulcerative Colitis (UC) and Crohn's Disease (CD) GWAS}

\subsubsection*{\textit{Tissue-level investigation using GenoSkyline (GS) annotations}}

We also applied the multi-GPA-Tree approach to the UC and CD GWAS and tissue-specific GS annotations to identify SNPs that were marginally and jointly associated with UC and CD, and to characterize the functional relevance of the single and multiple trait risk-associated SNPs. At the nominal global FDR level of $0.05$, multi-GPA-Tree identified $5,041$ SNPs that were jointly associated with both UC and CD, $5,430$ SNPs that were marginally associated with UC and $5,041$ SNPs that were marginally associated with CD (Table \ref{reftab1}). In this joint analysis, the original multi-GPA-Tree model identified blood tissue at the root node and included $2$ leaves (Fig. \ref{fig:tree_plot}C). Further investigation showed that $1,319$ SNPs that were jointly associated with both UC and CD, $1,453$ SNPs that were marginally associated with UC and $1,319$ SNPs that were marginally associated with CD were also annotated for blood tissue. Of the $1,319$ jointly associated and blood annotated SNPs, $990$ were protein-coding. Chromosomes $1$ and $2$ had the most number of protein-coding SNPs, followed by chromosomes $17$ and $5$. The \textit{THADA} and \textit{ATG16L1} genes in chromosome $2$, \textit{C5orf56} gene in chromosome $5$ and \textit{IKZF3} gene in chromosome $17$ had the most number of protein-coding SNPs with $9$ SNPs each related to coding the \textit{THADA} and \textit{IKZF3} genes, and $8$ SNPs each related to coding the \textit{ATG16L1} and \textit{C5orf56} genes. Although not directly implicated in the pathogenesis of UC or CD, the \textit{THADA} gene is known to influence metabolic mechanisms like adipogenesis \cite{pauct}. In contrast, genetic variants of the \textit{ATG16L1} gene are some of the most studied in the pathogenesis of Crohn's disease, playing a role in pathogen clearance, cytokine production, protein regulation and endoplasmic stress response control \cite{salem, hampe}. Similarly, increased expression of \textit{IKZF3}, a transcription factor that plays an important role in the regulation of B lymphocyte proliferation and differentiation, has been observed in patients with CD and UC \cite{huangc, soderman}, and \textit{C5orf56} is known to influence the immune stimulus specific enhancer for \textit{IRF1}, a gene established in the pathogenesis of Crohn's disease \cite{leonas, brandtm, huffcd}. 

We also discovered $7$ SNPs each in chromosomes $5, 6$ and $9$ known to code the \textit{FYB}, \textit{BACH2} and \textit{DOCK8} genes, and $6$ SNPs each known to code the \textit{BANK1}, \textit{LEF1}, and \textit{NFKB1} genes in chromosome $4$. The \textit{FYB} gene is related to T cells signaling and plays a role in IL-2A expression, and is known to be associated with some autoimmune regulation \cite{azevedo, addobbati}. Likewise, \textit{BACH2} is a critical gene for B cell regulatory activity and T cell function and differentiation and is a known susceptibility locus for CD and UC \cite{laffin, zhangb}; \textit{DOCK8} is known to regulate diverse immune sub-types including lymphocytes and plays a role in immune synapse formation and pathogen proliferation \cite{kearneycj}; and \textit{NFKB1} is a known transcription regulator of immune response, apoptosis and cell proliferation and is up-regulated in both UC and CD patients \cite{karban}. On the contrary, although \textit{BANK1} is a B cell gene known to be associated with SLE and RA \cite{orozco, kozyrev}, only few studies linking specific \textit{BANK1} variants to CD has been published \cite{lid, jostins} and it's role in the pathogenesis of both CD and UC remains understudied. This is also true for the \textit{LEF1} gene, a known mediator in the Wnt signaling pathway \cite{beisner}.\\

In the individual trait analysis for UC and CD using the GPA-Tree approach, we identified blood, GI and epithelium tissues for UC, and blood and epithelium tissues for CD with blood tissue at the root node for both traits. GPA-Tree identified $1,566$ SNPs to be associated with UC and $3,185$ SNPs to be associated with CD with $540$ UC associated and $960$ CD associated SNPs also annotated for blood tissue. Among the top $4$ genes identified in the joint analysis of UC, CD and GS annotations, one or more SNPs related to the \textit{THADA}, \textit{IKZF3} and \textit{C5orf56} genes were also identified in the single trait analysis of both UC and CD. However, SNPs related to \textit{ATG16L1} gene were identified for CD only.

\subsubsection*{\textit{Cell-type-level investigation using GenoSkylinePlus (GSP) annotations}}

In the second phase of the analysis, we combined the UC and CD GWAS and 10 blood related cell-type specific GSP annotations using the multi-GPA-Tree approach. At the nominal global FDR level of $0.05$, multi-GPA-Tree identified $4,576$ SNPs that were jointly associated with UC and CD, $4,995$ SNPs that were marginally associated with UC and $4,912$ SNPs that were marginally associated with CD (Table \ref{reftab1}). The original multi-GPA-Tree model fit identified primary natural killer cells at the root node and included $2$ leaves (Fig. \ref{fig:tree_plot}D). Further investigation showed that $507$ SNPs that were jointly associated with both UC and CD, $579$ SNPs that were marginally associated with UC and $554$ SNPs that were marginally associated with CD were annotated for natural killer cells. Of the $507$ jointly associated and natural killer cells annotated SNPs, $360$ were protein-coding. Chromosomes $1$ and $2$ had the most number of protein-coding SNPs, followed by chromosomes $5$ and $17$. The \textit{C5orf56} and \textit{IRF1} genes in chromosome $5$ and \textit{FAM53B} gene in chromosome $10$ had the most number of protein-coding SNPs with $8$ SNPs related to coding the \textit{C5orf56} gene, $5$ SNPs related to coding the \textit{IRF1} gene and $4$ SNPs related to coding the \textit{FAM53B} gene. We also discovered $3$ SNPs each known to code the \textit{ATG16L1} and \textit{THADA} genes in chromosome $2$, \textit{IKZF3} and \textit{PGAP3} genes in chromosome $17$, \textit{DOCK8} gene in chromosome $9$, \textit{TSPAN14} gene in chromosome $10$ and \textit{ETS1} gene in chromosome $11$. \textit{FAM53B} is known to be associated with humoral immune reponse, regulation of immune effector process, and regulation of lymphocyte activation \cite{xuq}; reduced expression of \textit{PGAP3} is known to be related to impaired clearance of apoptotic cells and has been observed in CD and UC patients \cite{soderman}; \textit{TSPAN14} is expressed in immune cell types participating in immunity and inflammation, and is positively correlated with microphages and neutrophils and negatively correlated with T cells CD8 \cite{liq2022}; and finally, \textit{ETS1} is known to be over-expressed in intestinal epithelial cells of patients with UC \cite{lil}, and has also been linked to fistula formation, an epithelial defect caused by destructive inflammation, in the pathogenesis of CD \cite{scharl}.\\

The individual analysis using the GPA-Tree approach identified primary natural killer, monocytes, effector/memory enriched T and regulatory T cells with natural killer cells at the root node for both UC and CD. Additionally, primary B cells was also identified for CD. In the individual GPA-Tree analysis, we identified $1,654$ SNPs to be associated with UC and $3,232$ SNPs to be associated with CD. Of those associated with UC, $186$ were annotated for both natural killer and effector/memory enriched T cells, $134$ were annotated for natural killer cells but not for effector/memory enriched T cells, $112$ were annotated for monocytes and $127$ were annotated for regulatory T cells. Similarly, of those associated with CD, $278$ were annotated for both natural killer and effector/memory enriched T cells, $211$ were annotated for natural killer cells but not for effector/memory enriched T cells, $181$ were annotated for monocytes,  $161$ were annotated for regulatory T and $132$ were annotated for primary B cells. Among the top 3 genes identified in the joint analysis of UC, CD and GSP annotations, one or more protein-coding SNPs related to the \textit{C5orf56}, \textit{IRF1} and \textit{FAM53B} genes were also identified in the single trait analysis of both UC and CD. 
%PLOS does not support heading levels beyond the 3rd (no 4th level headings).

\section*{Discussion}

Over the past 20 years, several GWAS have been conducted, leading to successful identification of over two hundred thousand trait risk-associated genetic variants \cite{buniello2019nhgri}. The advancement in complexity of newer statistical approaches to exploit the richness in GWAS data even further has been helpful in identifying many previously unknown genetic variants and it is expected that newer discoveries are forthcoming. Current findings have been crucial in identifying treatment therapies and for new drug discoveries \cite{shul, breeng, visscher}. Yet, a crucial gap that needs to be filled with new variant discovery is in our understanding of the functional mechanisms and pathways through which genetic variants influence traits. It is well known that complex traits are often caused by an amalgamation of functional mechanisms that can be described by multiple functional annotations \cite{petronis, zhangw}. Therefore, identifying the combinations of functional annotations that are associated with the traits can provide valuable insight into trait etiology. However, to the best of our knowledge, we are currently lacking statistical methodologies that identify the combinations of functional annotations that act in unison to influence traits. We propose the discussed multi-GPA-Tree approach to fill in this gap.

In comparison to existing methods, the overall strength of the multi-GPA-Tree approach is that it can automatically select the combinations of functional annotations from a group of annotations without excessively increasing the complexity of the model and be used to benefit our understanding of the functional mechanisms related to a single or multiple traits. The multi-GPA-Tree approach achieves that goal by following a hierarchical architecture that combines an iterative procedure (EM algorithm) and a multivariate decision tree algorithm. During simulation study, the multi-GPA-Tree approach showed consistently better performance than the LPM approach in terms of AUC, statistical power and type-I error control in identifying trait risk-associated variants for single and multiple traits and also distinctly identified relevant annotations from noise annotations with great accuracy (Fig \ref{fig2}).

In real data application, multi-GPA-Tree showed increased efficiency in identifying risk-associated SNPs for both traits when two traits are jointly analyzed and validated some annotational findings already established in literature (Table \ref{reftab1}, Fig \ref{fig:tree_plot}). We compared the real data application findings from multi-GPA-Tree to findings from our recently published method `GPA-Tree' \cite{gpatree}, a statistical approach that does not exploit the pleiotropic relationship between traits and prioritizes variants that are marginally associated with a single trait. Our comparison demonstrated that multi-GPA-Tree consistently identified more marginally risk-associated variants for both traits when the traits are analyzed jointly. Evidently, while GPA-Tree identified more annotations to be relevant with a specific trait, multi-GPA-Tree identified annotations that are largely common between the two jointly analyzed traits. For instance, blood tissue was identified in both joint and individual analysis of SLE and RA, and UC and CD. Similarly, regulatory T cells was identified as a relevant annotation when SLE and RA were jointly analyzed which was also a common annotation identified when SLE and RA are individually analyzed. Likewise, natural killer cells was identified as a relevant annotation when UC and CD were jointly analyzed, again a common annotation also identified for both UC and CD when individually analyzed. Overall, these results are consistent with previous literature indicating connections between autoimmune diseases like SLE, RA, UC and CD and blood tissue \cite{bloodsle1, bloodra1, blooduccd1}, and SLE and RA and regulatory T cells \cite{comte, ohl, hoffman, toh, leipe}, and UC and CD and natural killer cells \cite{manzano, takayama, poggi, auer}. Moreover, in addition to identifying a few candidate genes (\textit{RASA2, TXNDC11, THADA}) for SLE, RA, UC and CD that have previously been linked to other allergy, thyroid or metabolic traits, we also validated previous findings linking the \textit{PLCL1}, \textit{IL2RA} and \textit{UHRF1BP1} genes to SLE and RA \cite{wangq, zhour, kasher, carr, hinks, gateva, ramosps, delgado}, and the \textit{ATG16L1, C5orf56} and \textit{IKZF3} genes to UC and CD \cite{leescw, glasj, fishersa, leonas, cruzromero, brandtm, soderman, dinarzo}.

From the statistical modeling perspective, several assumptions are made in multi-GPA-Tree. First, we assume that the genetic variants are conditionally independent given its functional information which greatly simplifies our model and leads to efficient computation of the parameter estimates. Although this assumption omits the linkage disequilibrium (LD) structure present between SNPs in the same genomic region, it still allows us to conservatively infer risk-associated variants by modestly controlling the type-I error rate by potentially also identifying SNPs that are in LD with each other to be risk-associated. Second, we assume that signal in the GWAS association p-values are related to the functional potential of a SNP, so some functional signal should be present in the GWAS and annotation data for the multi-GPA-Tree approach to work efficiently. Simulation results suggest that at least $10\%$ of variants should be functionally annotated for at least one feature to achieve valid parameter estimates and controlled type-I error at the nominal level.

Here we have presented a novel statistical approach, named multi-GPA-Tree, that can exploit pleiotropic relationship between multiple GWAS data and integrate GWAS data and tissue and cell-type specific functional annotation data in an efficient manner. Compared to some existing approaches which require genotype data at the individual level and annotation data that follows certain distributional assumption, multi-GPA-Tree only requires summary statistics for GWAS data and binary annotation data for analysis. These features make multi-GPA-Tree an attractive and effective tool for the integrative analysis of GWAS data with functional annotation data. Despite the promising statistical improvements made by multi-GPA-Tree, the biological implications need to be independently replicated and newly identified variants need to be independently validated. Two limitations of multi-GPA-Tree are that it cannot handle continuous or count annotation data and scaling multi-GPA-Tree to more than two traits can still be computationally challenging. Addressing issues related to integrating multiple GWAS and multiple types of annotation data are important areas of our future work.

\section*{Acknowledgments}
We thank Dr. Paula S. Ramos, Dr. Andrew Lawson  and Dr. Kelly J. Hunt from the Medical University of South Carolina and Dr. Hang J. Kim from the University of Cincinnati for useful discussion related to the topic, and for their guidance and support in completing this work. This work has been supported through grant support from the National Institute of General Medical Sciences (R01 GM122078), National Institute on Drug Abuse (U01 DA045300), National Human Genome Research Institute (R21 HG012482), National Institute on Aging (U54 AG075931), and the Pelotonia Institute of Immuno-Oncology (PIIO). The content is solely the responsibility of the authors and does not
necessarily represent the official views of the funders.

\section*{Conflict of Interest} None declared.
% \vspace*{-12pt}

%\nolinenumbers


\begin{thebibliography}{10}

\bibitem{buniello2019nhgri}
Buniello A, MacArthur JA, Cerezo M, Harris LW, Hayhurst J, Malangone C, McMahon A, Morales J, Mountjoy E, Sollis E, Suveges D. 
\newblock The NHGRI-EBI GWAS Catalog of published genome-wide association studies, targeted arrays and summary statistics 2019. 
\newblock Nucleic acids research. 2019 Jan 8;47(D1):D1005-12.


\bibitem{manolio1}
Manolio TA, Collins FS, Cox NJ, Goldstein DB, Hindorff LA, Hunter DJ, McCarthy MI, Ramos EM, Cardon LR, Chakravarti A, Cho JH. 
\newblock Finding the missing heritability of complex diseases. 
\newblock Nature. 2009 Oct;461(7265):747-53.

\bibitem{leesh}
Lee SH, Wray NR, Goddard ME, Visscher PM. 
\newblock Estimating missing heritability for disease from genome-wide association studies. 
\newblock The American Journal of Human Genetics. 2011 Mar 11;88(3):294-305.

\bibitem{maherb}
Maher B.
\newblock Personal genomes: The case of the missing heritability. 
\newblock Nature. 2008 Nov 6;456(7218):18-22.

\bibitem{nikpay}
Nikpay M, Goel A, Won HH, Hall LM, Willenborg C, Kanoni S, Saleheen D, Kyriakou T, Nelson CP, Hopewell JC, Webb TR. 
\newblock A comprehensive 1000 Genomes-based genome-wide association meta-analysis of coronary artery disease. 
\newblock Nature Genetics. 2015;47(10):1121.

\bibitem{priceal}
Price AL, Spencer CC, Donnelly P. 
\newblock Progress and promise in understanding the genetic basis of common diseases. 
\newblock Proceedings of the Royal Society B: Biological Sciences. 2015 Dec 22;282(1821):20151684.

\bibitem{kundaje2015integrative}
Kundaje A, Meuleman W, Ernst J, Bilenky M, Yen A, Heravi-Moussavi A, Kheradpour P, Zhang Z, Wang J, Ziller MJ, Amin V. 
\newblock Integrative analysis of 111 reference human epigenomes. 
\newblock Nature. 2015 Feb;518(7539):317-30.


\bibitem{Andreassen}
Andreassen OA, Djurovic S, Thompson WK, Schork AJ, Kendler KS, O’Donovan MC, Rujescu D, Werge T, van de Bunt M, Morris AP, McCarthy MI. 
\newblock Improved detection of common variants associated with schizophrenia by leveraging pleiotropy with cardiovascular-disease risk factors. 
\newblock The American Journal of Human Genetics. 2013 Feb 7;92(2):197-209.

\bibitem{stearns}
Stearns FW. 
\newblock One hundred years of pleiotropy: a retrospective. 
\newblock Genetics. 2010 Nov 1;186(3):767-73.

\bibitem{chung2017graph}
Chung D, Kim HJ, Zhao H. 
\newblock{graph-GPA: a graphical model for prioritizing GWAS results and investigating pleiotropic architecture}. 
\newblock PLoS computational biology. 2017 Feb 17;13(2):e1005388.

\bibitem{dbgap}
Mailman MD, Feolo M, Jin Y, Kimura M, Tryka K, Bagoutdinov R, Hao L, Kiang A, Paschall J, Phan L, Popova N. 
\newblock The NCBI dbGaP database of genotypes and phenotypes. 
\newblock Nature genetics. 2007 Oct;39(10):1181-6.

\bibitem{giral2018into}
Giral H, Landmesser U, Kratzer A. 
\newblock{Into the wild: GWAS exploration of non-coding RNAs.}
\newblock Frontiers in cardiovascular medicine. 2018 Dec 17;5:181.

\bibitem{farh2015genetic}
Farh KK, Marson A, Zhu J, Kleinewietfeld M, Housley WJ, Beik S, Shoresh N, Whitton H, Ryan RJ, Shishkin AA, Hatan M. 
\newblock Genetic and epigenetic fine mapping of causal autoimmune disease variants. \newblock Nature. 2015 Feb;518(7539):337-43.

\bibitem{maurano2012systematic}
Maurano MT, Humbert R, Rynes E, Thurman RE, Haugen E, Wang H, Reynolds AP, Sandstrom R, Qu H, Brody J, Shafer A. 
\newblock{Systematic localization of common disease-associated variation in regulatory DNA.}
\newblock Science. 2012 Sep 7;337(6099):1190-5.

\bibitem{Schork2013a}
Schork AJ, Thompson WK, Pham P, Torkamani A, Roddey JC, Sullivan PF, Kelsoe JR, O'donovan MC, Furberg H, Tobacco and Genetics Consortium, Bipolar Disorder Psychiatric Genomics Consortium. 
\newblock All SNPs are not created equal: genome-wide association studies reveal a consistent pattern of enrichment among functionally annotated SNPs. 
\newblock PLoS genetics. 2013 Apr 25;9(4):e1003449.


\bibitem{Ming2018}
Ming J, Dai M, Cai M, Wan X, Liu J, Yang C. 
\newblock LSMM: a statistical approach to integrating functional annotations with genome-wide association studies. 
\newblock Bioinformatics. 2018 Aug 15;34(16):2788-96.


\bibitem{Zablocki2014}
Zablocki RW, Schork AJ, Levine RA, Andreassen OA, Dale AM, Thompson WK. 
\newblock Covariate-modulated local false discovery rate for genome-wide association studies. 
\newblock Bioinformatics. 2014 Aug 1;30(15):2098-104.

\bibitem{gtex2015genotype}
GTEx Consortium, Ardlie KG, Deluca DS, Segrè AV, Sullivan TJ, Young TR, Gelfand ET, Trowbridge CA, Maller JB, Tukiainen T, Lek M. 
\newblock The Genotype-Tissue Expression (GTEx) pilot analysis: multitissue gene regulation in humans. 
\newblock Science. 2015 May 8;348(6235):648-60.



\bibitem{chung2014gpa}
Chung D, Yang C, Li C, Gelernter J, Zhao H. 
\newblock GPA: a statistical approach to prioritizing GWAS results by integrating pleiotropy and annotation. 
\newblock PLoS genetics. 2014 Nov 13;10(11):e1004787.


\bibitem{Ming439133}
Ming J, Wang T, Yang C. 
\newblock LPM: a latent probit model to characterize the relationship among complex traits using summary statistics from multiple GWASs and functional annotations. 
\newblock Bioinformatics. 2020 Apr 15;36(8):2506-14.

\bibitem{gpatree}
Khatiwada A, Wolf BJ, Yilmaz AS, Ramos PS, Pietrzak M, Lawson A, Hunt KJ, Kim HJ, Chung D. 
\newblock GPA-Tree: statistical approach for functional-annotation-tree-guided prioritization of GWAS results. 
\newblock Bioinformatics. 2022 Feb 15;38(4):1067-74.

\bibitem{de2002multivariate}
De'Ath G. 
\newblock Multivariate regression trees: a new technique for modeling species–environment relationships.
\newblock Ecology. 2002 Apr;83(4):1105-17.

\bibitem{moontk}
Moon TK. 
\newblock The expectation-maximization algorithm. 
\newblock IEEE Signal processing magazine. 1996 Nov;13(6):47-60.

\bibitem{newton2004detecting}
Newton MA, Noueiry A, Sarkar D, Ahlquist P. 
\newblock Detecting differential gene expression with a semiparametric hierarchical mixture method. 
\newblock Biostatistics. 2004 Apr 1;5(2):155-76.


\bibitem{gohlke}
Gohlke JM, Thomas R, Zhang Y, Rosenstein MC, Davis AP, Murphy C, Becker KG, Mattingly CJ, Portier CJ. \newblock Genetic and environmental pathways to complex diseases. 
\newblock BMC Systems Biology. 2009 Dec;3(1):1-5.

\bibitem{kimya}
Kim YA, Wuchty S, Przytycka TM. 
\newblock Identifying causal genes and dysregulated pathways in complex diseases. 
\newblock PLoS computational biology. 2011 Mar 3;7(3):e1001095.


\bibitem{jia2020identification}
Jia X, Shi N, Feng Y, Li Y, Tan J, Xu F, Wang W, Sun C, Deng H, Yang Y, Shi X. 
\newblock {Identification of 67 pleiotropic genes associated with seven autoimmune$/$ autoinflammatory diseases using multivariate statistical analysis}. 
\newblock Frontiers in Immunology. 2020 Feb 3;11:30.

\bibitem{lee2019genomic}
Lee PH, Anttila V, Won H, Feng YC, Rosenthal J, Zhu Z, Tucker-Drob EM, Nivard MG, Grotzinger AD, Posthuma D, Wang MM. 
\newblock Genomic relationships, novel loci, and pleiotropic mechanisms across eight psychiatric disorders. \newblock Cell. 2019 Dec 12;179(7):1469-82

\bibitem{sivakumaran2011abundant}
Sivakumaran S, Agakov F, Theodoratou E, Prendergast JG, Zgaga L, Manolio T, Rudan I, McKeigue P, Wilson JF, Campbell H. 
\newblock Abundant pleiotropy in human complex diseases and traits. 
\newblock The American Journal of Human Genetics. 2011 Nov 11;89(5):607-18.


\bibitem{langefeld2017transancestral}
Langefeld CD, Ainsworth HC, Graham DS, Kelly JA, Comeau ME, Marion MC, Howard TD, Ramos PS, Croker JA, Morris DL, Sandling JK. 
\newblock Transancestral mapping and genetic load in systemic lupus erythematosus. 
\newblock Nature communications. 2017 Jul 17;8(1):1-8.

\bibitem{okada2014genetics}
Okada Y, Wu D, Trynka G, Raj T, Terao C, Ikari K, Kochi Y, Ohmura K, Suzuki A, Yoshida S, Graham RR. 
\newblock Genetics of rheumatoid arthritis contributes to biology and drug discovery. 
\newblock Nature. 2014 Feb;506(7488):376-81.

\bibitem{de2017genome}
De Lange KM, Moutsianas L, Lee JC, Lamb CA, Luo Y, Kennedy NA, Jostins L, Rice DL, Gutierrez-Achury J, Ji SG, Heap G. 
\newblock Genome-wide association study implicates immune activation of multiple integrin genes in inflammatory bowel disease. 
\newblock Nature genetics. 2017 Feb;49(2):256-61.

\bibitem{lu2016integrative}
Lu Q, Powles RL, Wang Q, He BJ, Zhao H. 
\newblock Integrative tissue-specific functional annotations in the human genome provide novel insights on many complex traits and improve signal prioritization in genome wide association studies. 
\newblock PLoS genetics. 2016 Apr 8;12(4):e1005947.

\bibitem{lu2017systematic}
Lu Q, Powles RL, Abdallah S, Ou D, Wang Q, Hu Y, Lu Y, Liu W, Li B, Mukherjee S, Crane PK. 
\newblock Systematic tissue-specific functional annotation of the human genome highlights immune-related DNA elements for late-onset Alzheimer’s disease. 
\newblock PLoS genetics. 2017 Jul 24;13(7):e1006933.


\bibitem{luos}
Luo S, Li XF, Yang YL, Song B, Wu S, Niu XN, Wu YY, Shi W, Huang C, Li J. 
\newblock PLCL1 regulates fibroblast-like synoviocytes inflammation via NLRP3 inflammasomes in rheumatoid arthritis. 
\newblock Advances in Rheumatology. 2022 Jul 22;62.

\bibitem{carrej}
Carr EJ, Clatworthy MR, Lowe CE, Todd JA, Wong A, Vyse TJ, Kamesh L, Watts RA, Lyons PA, Smith KG. 
\newblock Contrasting genetic association of IL2RA with SLE and ANCA–associated vasculitis. 
\newblock BMC Medical Genetics. 2009 Dec;10(1):1-7.

\bibitem{caruso}
Caruso C, Candore G, Cigna D, Colucci AT, Modica MA. 
\newblock Biological significance of soluble IL-2 receptor. 
\newblock Mediators of inflammation. 1993 Jan 1;2(1):3-21.

\bibitem{gateva}
Gateva V, Sandling JK, Hom G, Taylor KE, Chung SA, Sun X, Ortmann W, Kosoy R, Ferreira RC, Nordmark G, Gunnarsson I. 
\newblock A large-scale replication study identifies TNIP1, PRDM1, JAZF1, UHRF1BP1 and IL10 as risk loci for systemic lupus erythematosus. 
\newblock Nature genetics. 2009 Nov;41(11):1228-33.

\bibitem{kozyrev}
Kozyrev SV, Abelson AK, Wojcik J, Zaghlool A, Reddy L, Prasad MV, Sanchez E, Gunnarsson I, Svenungsson E, Sturfelt G, Jönsen A. 
\newblock Functional variants in the B-cell gene BANK1 are associated with systemic lupus erythematosus. 
\newblock Nature genetics. 2008 Feb;40(2):211-6.

\bibitem{orozco}
Orozco G, Abelson AK, González‐Gay MA, Balsa A, Pascual‐Salcedo D, García A, Fernández‐Gutierrez B, Petersson I, Pons‐Estel B, Eimon A, Paira S. 
\newblock Study of functional variants of the BANK1 gene in rheumatoid arthritis. 
\newblock Arthritis \& Rheumatism: Official Journal of the American College of Rheumatology. 2009 Feb;60(2):372-9.

\bibitem{wangy}
Wang Y, Murakami Y, Yasui T, Wakana S, Kikutani H, Kinoshita T, Maeda Y. 
\newblock Significance of glycosylphosphatidylinositol-anchored protein enrichment in lipid rafts for the control of autoimmunity. 
\newblock Journal of Biological Chemistry. 2013 Aug 30;288(35):25490-9.

\bibitem{bowesj}
Bowes J, Ho P, Flynn E, Ali F, Marzo-Ortega H, Coates LC, Warren RB, McManus R, Ryan AW, Kane D, Korendowych E. 
\newblock Comprehensive assessment of rheumatoid arthritis susceptibility loci in a large psoriatic arthritis cohort. 
\newblock Annals of the rheumatic diseases. 2012 Aug 1;71(8):1350-4.

\bibitem{ferreira}
Ferreira MA, Vonk JM, Baurecht H, Marenholz I, Tian C, Hoffman JD, Helmer Q, Tillander A, Ullemar V, Van Dongen J, Lu Y. 
\newblock Shared genetic origin of asthma, hay fever and eczema elucidates allergic disease biology. 
\newblock Nature genetics. 2017 Dec;49(12):1752-7.

\bibitem{jaeger}
Jaeger M, Sloot YJ, Horst RT, Chu X, Koenen HJ, Koeken VA, Moorlag SJ, de Bree CJ, Mourits VP, Lemmers H, Dijkstra H. 
\newblock Thyrotrophin and thyroxine support immune homeostasis in humans. 
\newblock Immunology. 2021 Jun;163(2):155-68.

\bibitem{pauct}
Pau CT, Mosbruger T, Saxena R, Welt CK. 
\newblock Phenotype and tissue expression as a function of genetic risk in polycystic ovary syndrome. 
\newblock PloS one. 2017 Jan 9;12(1):e0168870.

\bibitem{salem}
Salem M, Ammitzboell M, Nys K, Seidelin JB, Nielsen OH. 
\newblock ATG16L1: a multifunctional susceptibility factor in Crohn disease. 
\newblock Autophagy. 2015 Apr 3;11(4):585-94.

\bibitem{hampe}
Hampe J, Franke A, Rosenstiel P, Till A, Teuber M, Huse K, Albrecht M, Mayr G, De La Vega FM, Briggs J, Günther S. 
\newblock A genome-wide association scan of nonsynonymous SNPs identifies a susceptibility variant for Crohn disease in ATG16L1. 
\newblock Nature genetics. 2007 Feb;39(2):207-11.

\bibitem{huangc}
Huang C, Haritunians T, Okou DT, Cutler DJ, Zwick ME, Taylor KD, Datta LW, Maranville JC, Liu Z, Ellis S, Chopra P. 
\newblock Characterization of genetic loci that affect susceptibility to inflammatory bowel diseases in African Americans. 
\newblock Gastroenterology. 2015 Nov 1;149(6):1575-86.

\bibitem{soderman}
Söderman J, Berglind L, Almer S. 
\newblock Gene expression-genotype analysis implicates GSDMA, GSDMB, and LRRC3C as contributors to inflammatory bowel disease susceptibility. 
\newblock BioMed research international. 2015 Oct;2015.

\bibitem{leonas}
León AS, Bernstein CN, El-Gabalawy H, Eck P. 
\newblock Variations in the IBD5 locus confer the risk of inflammatory bowel disease in a Manitoban Caucasian Cohort. 
\newblock Clin Nutr. 2018;5:1-6.

\bibitem{brandtm}
Brandt M, Kim-Hellmuth S, Ziosi M, Gokden A, Wolman A, Lam N, Recinos Y, Daniloski Z, Morris JA, Hornung V, Schumacher J. 
\newblock An autoimmune disease risk variant: A trans master regulatory effect mediated by IRF1 under immune stimulation?
\newblock PLoS genetics. 2021 Jul 27;17(7):e1009684.

\bibitem{huffcd}
Huff CD, Witherspoon DJ, Zhang Y, Gatenbee C, Denson LA, Kugathasan S, Hakonarson H, Whiting A, Davis CT, Wu W, Xing J. 
\newblock Crohn’s disease and genetic hitchhiking at IBD5. 
\newblock Molecular biology and evolution. 2012 Jan 1;29(1):101-11.

\bibitem{azevedo}
Azevedo Silva JD, Addobbati C, Sandrin-Garcia P, Crovella S. 
\newblock Systemic lupus erythematosus: old and new susceptibility genes versus clinical manifestations. 
\newblock Current Genomics. 2014 Feb 1;15(1):52-65.

\bibitem{addobbati}
Addobbati C, Brandão LA, Guimarães RL, Pancotto JA, Donadi EA, Crovella S, Segat L, Sandrin-Garcia P. 
\newblock FYB gene polymorphisms are associated with susceptibility for systemic lupus erythemathosus (SLE). 
\newblock Human immunology. 2013 Aug 1;74(8):1009-14.

\bibitem{laffin}
Laffin MR, Fedorak RN, Wine E, Dicken B, Madsen KL. 
\newblock A BACH2 gene variant is associated with postoperative recurrence of Crohn's disease. 
\newblock Journal of the American College of Surgeons. 2018 May 1;226(5):902-8.

\bibitem{zhangb}
Zhang B, Sun T. 
\newblock Transcription factors that regulate the pathogenesis of ulcerative colitis. \newblock BioMed Research International. 2020 Aug 24;2020.

\bibitem{kearneycj}
Kearney CJ, Randall KL, Oliaro J. 
\newblock DOCK8 regulates signal transduction events to control immunity. 
\newblock Cellular \& molecular immunology. 2017 May;14(5):406-11.

\bibitem{karban}
Karban AS, Okazaki T, Panhuysen CI, Gallegos T, Potter JJ, Bailey-Wilson JE, Silverberg MS, Duerr RH, Cho JH, Gregersen PK, Wu Y. 
\newblock Functional annotation of a novel NFKB1 promoter polymorphism that increases risk for ulcerative colitis. 
\newblock Human molecular genetics. 2004 Jan 1;13(1):35-45.

\bibitem{lid}
Li D, Achkar JP, Haritunians T, Jacobs JP, Hui KY, D'Amato M, Brand S, Radford-Smith G, Halfvarson J, Niess JH, Kugathasan S. 
\newblock A pleiotropic missense variant in SLC39A8 is associated with Crohn’s disease and human gut microbiome composition. 
\newblock Gastroenterology. 2016 Oct 1;151(4):724-32.

\bibitem{jostins}
Jostins L, Ripke S, Weersma RK, Duerr RH, McGovern DP, Hui KY, Lee JC, Philip Schumm L, Sharma Y, Anderson CA, Essers J. 
\newblock Host–microbe interactions have shaped the genetic architecture of inflammatory bowel disease. 
\newblock Nature. 2012 Nov;491(7422):119-24.

\bibitem{beisner}
Beisner J, Teltschik Z, Ostaff MJ, Tiemessen MM, Staal FJ, Wang G, Gersemann M, Perminow G, Vatn MH, Schwab M, Stange EF. 
\newblock TCF-1-mediated Wnt signaling regulates Paneth cell innate immune defense effectors HD-5 and-6: implications for Crohn's disease. 
\newblock American Journal of Physiology-Gastrointestinal and Liver Physiology. 2014 Sep 1;307(5):G487-98.

\bibitem{xuq}
Xu Q, Chen S, Hu Y, Huang W. 
\newblock Clinical M2 macrophages-related genes to aid therapy in pancreatic ductal adenocarcinoma. 
\newblock Cancer cell international. 2021 Dec;21(1):1-7.

\bibitem{liq2022}
Li Q, Gao X, Luo X, Wu Q, He J, Liu Y, Xue Y, Wu S, Rao F. 
\newblock Identification of Hub Genes Associated with Immune Infiltration in Cardioembolic Stroke by Whole Blood Transcriptome Analysis. 
\newblock Disease Markers. 2022 Jan 15;2022.

\bibitem{lil}
Li L, Miao X, Ni R, Miao X, Wang L, Gu X, Yan L, Tang Q, Zhang D. 
\newblock Epithelial-specific ETS-1 (ESE1$\/$ELF3) regulates apoptosis of intestinal epithelial cells in ulcerative colitis via accelerating NF$-\kappa$B activation. 
\newblock Immunologic research. 2015 Jun;62(2):198-212.

\bibitem{scharl}
Scharl M, Rogler G. 
\newblock Pathophysiology of fistula formation in Crohn's disease. 
\newblock World journal of gastrointestinal pathophysiology. 2014 Aug 8;5(3):205.



\bibitem{shul}
Shu L, Blencowe M, Yang X. 
\newblock Translating GWAS findings to novel therapeutic targets for coronary artery disease. 
\newblock Frontiers in cardiovascular medicine. 2018 May 30;5:56.

\bibitem{breeng}
Breen G, Li Q, Roth BL, O'donnell P, Didriksen M, Dolmetsch R, O'reilly PF, Gaspar HA, Manji H, Huebel C, Kelsoe JR. 
\newblock Translating genome-wide association findings into new therapeutics for psychiatry. 
\newblock Nature neuroscience. 2016 Nov;19(11):1392-6.

\bibitem{visscher}
Visscher PM, Brown MA, McCarthy MI, Yang J. 
\newblock Five years of GWAS discovery. 
\newblock The American Journal of Human Genetics. 2012 Jan 13;90(1):7-24.

\bibitem{petronis}
Petronis A. 
\newblock Epigenetics as a unifying principle in the aetiology of complex traits and diseases. 
\newblock Nature. 2010 Jun;465(7299):721-7.

\bibitem{zhangw}
Zhang W, Voloudakis G, Rajagopal VM, Readhead B, Dudley JT, Schadt EE, Björkegren JL, Kim Y, Fullard JF, Hoffman GE, Roussos P. 
\newblock Integrative transcriptome imputation reveals tissue-specific and shared biological mechanisms mediating susceptibility to complex traits. 
\newblock Nature communications. 2019 Aug 23;10(1):1-3.

\bibitem{bloodsle1}
Wahren-Herlenius M, Dörner T. 
\newblock Immunopathogenic mechanisms of systemic autoimmune disease. 
\newblock The Lancet. 2013 Aug 31;382(9894):819-31.

\bibitem{bloodra1}
Smith JB, Haynes MK. 
\newblock Rheumatoid arthritis—a molecular understanding. 
\newblock Annals of internal medicine. 2002 Jun 18;136(12):908-22.

\bibitem{blooduccd1}
Gleeson MH, Walker JS, Wentzel J, Chapman JA, Harris R. 
\newblock Human leucocyte antigens in Crohn's disease and ulcerative colitis. 
\newblock Gut. 1972 Jun 1;13(6):438-40.

\bibitem{comte}
Comte D, Karampetsou MP, Tsokos GC. 
\newblock T cells as a therapeutic target in SLE. 
\newblock Lupus. 2015 Apr;24(4-5):351-63.

\bibitem{ohl}
Ohl K, Tenbrock K. 
\newblock Regulatory T cells in systemic lupus erythematosus. 
\newblock European journal of immunology. 2015 Feb;45(2):344-55.

\bibitem{hoffman}
Hoffman RW. 
\newblock T cells in the pathogenesis of systemic lupus erythematosus. 
\newblock Clinical Immunology. 2004 Oct 1;113(1):4-13.

\bibitem{toh}
Toh ML, Miossec P. 
\newblock The role of T cells in rheumatoid arthritis: new subsets and new targets. \newblock Current opinion in rheumatology. 2007 May 1;19(3):284-8.

\bibitem{leipe}
Leipe J, Skapenko A, Lipsky PE, Schulze-Koops H. 
\newblock Regulatory T cells in rheumatoid arthritis. 
\newblock Arthritis research \& therapy. 2005 Mar;7(3):1-7.

\bibitem{manzano}
Manzano L, Alvarez-Mon M, Abreu L, Vargas JA, De la Morena E, Corugedo F, Duràntez A. 
\newblock Functional impairment of natural killer cells in active ulcerative colitis: reversion of the defective natural killer activity by interleukin 2. 
\newblock Gut. 1992 Feb 1;33(2):246-51.

\bibitem{takayama}
Takayama T, Kamada N, Chinen H, Okamoto S, Kitazume MT, Chang J, Matuzaki Y, Suzuki S, Sugita A, Koganei K, Hisamatsu T. 
\newblock Imbalance of NKp44+ NKp46$-$ and NKp44$-$ NKp46$+$ natural killer cells in the intestinal mucosa of patients with Crohn's disease. 
\newblock Gastroenterology. 2010 Sep 1;139(3):882-92.

\bibitem{poggi}
Poggi A, Benelli R, Venè R, Costa D, Ferrari N, Tosetti F, Zocchi MR. 
\newblock Human gut-associated natural killer cells in health and disease. 
\newblock Frontiers in immunology. 2019 May 3;10:961.

\bibitem{auer}
Auer IO, Ziemer E, Sommer H. 
\newblock Immune status in Crohn's disease. V. Decreased in vitro natural killer cell activity in peripheral blood. 
\newblock Clinical and experimental immunology. 1980 Oct;42(1):41.

\bibitem{wangq}
Ramos PS, Criswell LA, Moser KL, Comeau ME, Williams AH, Pajewski NM, Chung SA, Graham RR, Zidovetzki R, Kelly JA, Kaufman KM. 
\newblock A comprehensive analysis of shared loci between systemic lupus erythematosus (SLE) and sixteen autoimmune diseases reveals limited genetic overlap. \newblock PLoS genetics. 2011 Dec 8;7(12):e1002406.

\bibitem{zhour}
Zhou R, Lin X, Li DY, Wang XF, Greenbaum J, Chen YC, Zeng CP, Lu JM, Ao ZX, Peng LP, Bai XC. 
\newblock Identification of novel genetic loci for osteoporosis and/or rheumatoid arthritis using cFDR approach. 
\newblock PLoS One. 2017 Aug 30;12(8):e0183842.

\bibitem{kasher}
Kasher M, Freidin MB, Williams FM, Cherny SS, Malkin I, Livshits G. 
\newblock Shared genetic architecture between rheumatoid arthritis and varying osteoporotic phenotypes. 
\newblock Journal of Bone and Mineral Research. 2022 Mar;37(3):440-53.

\bibitem{hinks}
Hinks A, Ke X, Barton A, Eyre S, Bowes J, Worthington J, UK Rheumatoid Arthritis Genetics Consortium, British Society of Paediatric and Adolescent Rheumatology Study Group, Thompson SD, Langefeld CD, Glass DN. 
\newblock Association of the IL2RA$\backslash$backslash CD25 gene with juvenile idiopathic arthritis. 
\newblock Arthritis \& Rheumatism. 2009 Jan;60(1):251-7.

\bibitem{carr}
Carr EJ, Clatworthy MR, Lowe CE, Todd JA, Wong A, Vyse TJ, Kamesh L, Watts RA, Lyons PA, Smith KG. 
\newblock Contrasting genetic association of IL2RAwith SLE and ANCA–associated vasculitis. 
\newblock BMC Medical Genetics. 2009 Dec;10(1):1-7.


\bibitem{ramosps}
Ramos PS, Shaftman SR, Ward RC, Langefeld CD. 
\newblock Genes associated with SLE are targets of recent positive selection. \newblock Autoimmune diseases. 2014 Oct;2014.

\bibitem{delgado}
Delgado-Vega A, Sánchez E, Löfgren S, Castillejo-López C, Alarcón-Riquelme ME. \newblock Recent findings on genetics of systemic autoimmune diseases. 
\newblock Current opinion in immunology. 2010 Dec 1;22(6):698-705.

\bibitem{leescw}
Lees CW, Barrett JC, Parkes M, Satsangi J. 
\newblock New IBD genetics: common pathways with other diseases. 
\newblock Gut. 2011 Dec 1;60(12):1739-53.

\bibitem{glasj}
Glas J, Konrad A, Schmechel S, Dambacher J, Seiderer J, Schroff F, Wetzke M, Roeske D, Török HP, Tonenchi L, Pfennig S. 
\newblock The ATG16L1 gene variants rs2241879 and rs2241880 (T300A) are strongly associated with susceptibility to Crohn's disease in the German population. 
\newblock Official journal of the American College of Gastroenterology| ACG. 2008 Mar 1;103(3):682-91.

\bibitem{fishersa}
Fisher SA, Tremelling M, Anderson CA, Gwilliam R, Bumpstead S, Prescott NJ, Nimmo ER, Massey D, Berzuini C, Johnson C, Barrett JC. 
\newblock Genetic determinants of ulcerative colitis include the ECM1 locus and five loci implicated in Crohn's disease. 
\newblock Nature genetics. 2008 Jun;40(6):710-2.

\bibitem{cruzromero}
Cruz-Romero C, Guo A, Bradley WF, Vicentini JR, Yajnik V, Gee MS. 
\newblock Novel Associations Between Genome-Wide Single Nucleotide Polymorphisms and MR Enterography Features in Crohn's Disease Patients. 
\newblock Journal of Magnetic Resonance Imaging. 2021 Jan;53(1):132-8.


\bibitem{dinarzo}
Di Narzo AF, Peters LA, Argmann C, Stojmirovic A, Perrigoue J, Li K, Telesco S, Kidd B, Walker J, Dudley J, Cho J. 
\newblock Blood and intestine eQTLs from an anti-TNF-resistant Crohn's disease cohort inform IBD genetic association loci. 
\newblock Clinical and translational gastroenterology. 2016 Jun;7(6):e177.

\end{thebibliography}
\end{document}